%% file: main.tex
\documentclass[preprint,12pt]{elsarticle}

\usepackage{graphicx}
\usepackage{subcaption}
\usepackage{float}
\usepackage{amssymb}
\usepackage{amsmath}

\usepackage{booktabs}
\usepackage{verbatim}
\usepackage{adjustbox}

\usepackage{xcolor}
\usepackage{caption}
\journal{Powder Technology}
\begin{document}
\begin{frontmatter}
%
\title{The influence of material properties and process parameters on the spreading process in additive manufacturing}
%
%
\author[msm,dpm]{Mohamad Yousef Shaheen\corref{cor1}}
\ead{m.y.shaheen@utwente.nl}

\author[msm]{Anthony R. Thornton}
\ead{a.r.thornton@utwente.nl}
\author[msm]{Stefan Luding}
\ead{s.luding@utwente.nl}
\author[msm]{Thomas Weinhart}
\ead{t.weinhart@utwente.nl}

\cortext[cor1]{Corresponding author.}

\address[msm]{Multi-Scale Mechanics, Faculty of Engineering Technology, University of Twente, Drienerlolaan 5, 7522 NB Enschede, NL}
\address[dpm]{Design, Production and Management, Faculty of Engineering Technology, University of Twente, Drienerlolaan 5, 7522 NB Enschede, NL}
\begin{abstract}
\input{abstract.tex}
\end{abstract}


\end{frontmatter}

%
\section{Introduction}
\input{intro.tex}
\section{Methods} \label{sec:methods}
\input{methods.tex}

\section{Results and discussion}\label{sec:results}
\input{results.tex}
\section{Conclusions}\label{sec:conclusions}
\input{conclusions.tex}

\section*{Acknowledgment}
This work was financially supported by NWO-TTW project 15050 ‘Multiscale modelling of agglomeration: Application to tabletting and selective laser sintering’
%
%

%

\bibliographystyle{model1-num-names.bst}
\bibliography{ref.bib}


%
\input{tables.tex}
\input{figures.tex}
\end{document}

%% file: abstract.tex
Laser powder bed fusion (LPBF) is an additive manufacturing (AM) technology. To achieve high product quality, the powder is best spread as a uniform, dense layer. The challenge for LPBF manufacturers is to develop a spreading process that can produce a consistent layer quality for the many powders used, which show considerable differences in spreadability. Therefore, we investigate the influence of material properties, process parameters and the type of spreading tool on the layer quality. The discrete particle method is used to simulate the spreading process and to define metrics to evaluate the powder layer characteristics. We found that particle shape and surface roughness in terms of rolling resistance and interparticle sliding friction as well as particle cohesion all have a major (sometimes surprising) influence on the powder layer quality: more irregular shaped particles, rougher particle surfaces and/or higher interfacial cohesion usually, but not always, lead to worse spreadability. Our findings illustrate that there is a trade-off between material properties and process parameters. Increasing the spreading speed decreases layer quality for non- and weakly cohesive powders, but improves it for strongly cohesive ones. Using a counter-clockwise rotating roller as a spreading tool improves the powder layer quality compared to spreading with a blade. Finally, for both geometries, a unique correlation between the quality criteria uniformity and mass fraction is reported and some of the findings are related to size-segregation during spreading.

%% file: intro.tex
Laser powder bed fusion (LPBF) is an additive manufacturing (AM) technology. In contrast to subtractive or formative methods, objects are produced from three-dimensional digital models in a layer-by-layer fashion. It offers design flexibility and easy customisation that contributed to its rapid growth and wide utilisation in different industrial sectors \cite{intro1,campbell2012,sames2016}. Fig.\,\ref{schematic} shows a schematic of the process. Parts are produced by spreading successive layers of powder material and solidifying selected parts by partially or fully melting them with a laser \cite{campbell2012,Kruth2007}.

The powder spreading process is governed by the geometry, speed, and material properties of the spreading tool. In addition, powder feedstock and powder characteristics play a major role for the powder layer quality, which in turn influences the final product properties and quality \cite{TAN2017,report,hebert2016,Vock2019}. 

The discrete particle method (DPM) has been recently used to simulate the spreading process in AM. Despite its computational expense, it is a powerful tool for simulating granular materials and understanding particulate system phenomena that are either inaccessible or difficult to obtain from experiments.

Early studies of Herbold et al. \cite{Herbold2015}, Mindt et al. \cite{Mindt2016} and Parteli et al. \cite{Parteli2016} have used DPM to simulate the spreading process in LPBF.
For example, Mindt et al. \cite{Mindt2016} have investigated the influence of the blade gap height: the distance between the powder bed or base plate and the spreading blade, on the spread layer of spherical Ti-6Al-4V powder. They concluded that a blade gap height equal to or less than the maximum particle diameter would result in a reduced packing density.
While Parteli et al. \cite{Parteli2016} simulated the spreading of PA12 powder using a multisphere method to model complex particle shapes with a counter-clockwise (cc) rotating roller as a spreading tool. They showed that the powder bed surface roughness is increased for a higher spreading speed and for powders with a wider size distribution.  
Parteli et al. \cite{Parteli2016} findings were confirmed by Haeri et al. \cite{Haeri2017}, who used DPM to spread rod-like particles. They found a higher surface roughness and smaller volume fractions for increased spreading speed and particle aspect ratios. In addition, they showed that a better layer quality is obtained by using a cc rotating roller instead of a blade. Another study by Haeri \cite{Haeri2017-2} investigated the optimization of the blade spreader geometry and found that using a super-elliptic edge profile would result in a better layer quality than a normal flat edge blade.

The flowability behaviour of spherical 316 L stainless steel particles during the spreading process was investigated by Chen et al. \cite{Chen2017}. They used DPM to simulate the process with a blade as a spreading tool. They found that decreasing either sliding or rolling friction would decrease the dynamic repose angle and thus improve flowability.
While Nan et al. \cite{Nan2018a} have reconstructed complex particle shape of 316 L stainless steel powder as a function of particle size. They investigated the period and frequency of transient jamming in powder spreading with a small gap height; they found relationships between particle properties, blade speed and gap height. Later, they studied 316 L stainless steel powder flow \cite{Nan2018b}. They showed that the mass flow rate through the gap, initially, increases linearly with the gap height until it reaches a limit beyond which the mass flow rate cannot be further increased.

Other studies have investigated particle cohesion influence on the spreading process \cite{Herbold2015,Meier2019a}. For instance, Meier et al. \cite{Meier2019a} introduced a DPM model for cohesion and were able to predict the effective surface energy of Ti-6Al-4V. Then they performed a parametric study, highlighting the effect of cohesion on the spreading process. They found that powder layer quality decreases as particle size decreases i.e. cohesiveness increases \cite{Meier2019b}.
Later, Han et al. \cite{Han2019} adapted the approach of Meier et al. \cite{Meier2019a} to calibrate the surface energy of Hastealloy X (HX) alloy. They found that a layer thickness of 40 $\mathrm{\mu m}$ produce a uniform powder bed spreading.

Recently, Chen et al. \cite{Chen2019} investigated 316L stainless steel powder layer packing density using a blade as a spreading tool. They found that there is a ``stress-dip" region at the bottom front of the spreader, and identified three mechanisms that affect the packing density of the powder layer: (1) The ``cohesion effect" causes particle agglomerates, (2) the ``wall effect" creates vacancies in the powder layer and (3) the ``percolation effect" leads to particle segregation.
While Fouda et al. \cite{Fouda2020} performed a DPM simulation of an idealized system with mono-sized particles using a blade as a spreading tool. They showed that the powder layer packing fraction is always lower than the initial powder heap due to three mechanisms, shear-induced dilation, particle rearrangement and particle inertia.

The ``spreadability" of a powder can be defined as the powder ability to spread under certain conditions to form a uniform and highly packed powder layer. Bad spreadability can lead to powder bed defects, segregation, non-uniform density and/or a loose particle packing, all of which have negative effects on the quality of the final product. 
Unfortunately, powders used in AM tend to behave differently under different conditions.
In addition, the recycling of powder changes the material properties, both chemical and morphological ones. Powder properties also depend on powder storage, contamination and environmental effects \cite{TAN2017,hebert2016,malekipour2018,DOWLING2020}.
Spherical particle shapes are favourable in terms of flowability and powder bed packing density \cite{Vock2019}. Fig.\,\ref{fig:Ti15} shows an SEM image of Ti-6Al-4V powder (produced by plasma rotating electrode) with spherical particles. However, non-sphericity is usually present due to satellites, fractured, adhered particles, etc, as shown in Fig.\,\ref{fig:Ti15defects}. Fig.\,\ref{fig:Ti45} shows an example of a Ti-6Al-4V powder (produced by gas atomization) with satellites and wider particle size distribution.\footnote{Due to the larger particles size distribution, this powder is usually used in electron beam melting (EBM) which is another AM technology that requires powder spreading}
The influence of the material properties and spreading process parameters on the spreadability have not been investigated enough in the literature. More specifically, the relationship between particle's shape, surface roughness, cohesiveness and process parameters has not been investigated yet.

In this work we perform a study of the influence of material and process parameters on the spreading process of a Ti-6Al-4V powder, using DPM simulations. The focus is on three material parameters: (i) the interparticle sliding friction as a measure of surface roughness: we assume that an increase in surface roughness causes a reduction in contact area, and thus an increased normal pressure, which causes plastic deformation of the asperities and thus sliding friction.
Fig.\,\ref{fig:sliding} shows a schematics of the contact area between two rough solid surfaces. The apparent area $A_a$ is much smaller than the actual contact area $A_r$, where only the highest asperities are in contact \cite{GMBook}.
(ii) Rolling resistance as a simple measure to mimic particle's shape, as an approximation of the behaviour of aspherical particles (resembling small asperities) \cite{GMBook,Wensrich2012,Wensrich2014}. Fig.\,\ref{fig:rolling} illustrates how rolling resistance results from the imbalance of the normal reaction force $f_{n}^{R}$ at the contact area when an external torque is applied. For example, the rolling behaviour of a polygonal particle can be modelled by a spherical particle with coefficient of rolling friction $\mu_r = \frac{c}{2 R}$, Fig.\,\ref{fig:rolling}. Wensrich et al. \cite{Wensrich2014} have demonstrated that a complex particle shape can be captured by rolling friction coefficient of a spherical particle, where ``a value of around half of the normalised average eccentricity (equivalent rolling friction)" was considered as an appropriate amount to capture the effect of particle shape.
(iii) effective dry cohesion, as determined by the interfacial surface energy.

The process parameters investigated are the spreading speed and tool geometry. The effect of the gap height and the layer thickness are not considered in this study. A small gap height, and thus a thinner powder layer is usually preferable to achieve higher resolution, i.e., better adhesion between constitutive powder layers during the sintering/melting process.
Our quantitative parametric study can provide a guidance for the calibration of DPM, showing trends and importance of material and process parameters. A calibration of the model parameters according to a specific powder material is not performed here. 
Fig.\,\ref{fig:simFramework} shows a flow chart of DPM simulation, calibration and validation framework.

Thus, this paper aims to answer the following questions (i)\textit{how to quantify the powder layer quality?}, (ii) \textit{what is the relation between those material properties and process parameters such as spreading speed or geometry?}, (iii) \textit{what is the effect of particle shape, roughness and cohesiveness on the spread powder layer for different process parameters?}

The remainder of this paper is divided into three sections. In section \ref{sec:methods}, we introduce the methods used, e.g. the DPM force law, the simulation setup, etc. We present and discuss the results in section \ref{sec:results}, before we conclude in section \ref{sec:conclusions}.

%% file: methods.tex
In section \ref{sec:dpmModel}, we introduce the force law used in the DPM. The exact DPM parameters used are detailed in section \ref{sec:modelParameters}. Then we describe the simulation setup in section \ref{sec:simSetup}. In section \ref{sec:DOS}, we illustrate the design of the parametric study. We define metrics to characterise the powder layer in section \ref{sec:layerMetrics}.
\subsection{Discrete particle method}\label{sec:dpmModel}
The discrete particle method is used to simulate the spreading process.
The interaction of $N$ poly-disperse particles is modeled using the standard linear spring-dashpot model \cite{Cundall} for the normal force. The normal force (parallel to $\mathbf{r}_{ij}$) is composed of a linear elastic, linear dissipative and a linear adhesive force:
\begin{equation}
f_{ij}^{n} = k_n \delta_{ij}^{n} + \eta_n \dot{\delta}_{ij}^{n} + f_{ij}^{adh},
\end{equation}
with a normal spring stiffness $k_n$, damping coefficient $\eta_n$, normal relative velocity $\dot{\delta}_{ij}^{n}$ and a linear adhesion force $f_{ij}^{adh}$. 
Each pair of particles $i$ and $j$ are in contact, if their overlap $\delta_{ij}^{n}$  is positive.
In addition, particles can interact with the base or the powder bed and the spreading tool, which is constructed from polygonal shapes.

Many models exist in DPM to describe dry cohesion of small particles, the attractive force due to van der Waals interaction between particles close to each other or in contact. For simplicity, a linear elastic adhesive force law (acting opposite to the normal elastic repulsive force) is used:
\begin{equation}
f_{ij}^{adh} =
  \begin{cases}
     -f_{max}^{adh} & \quad \delta_{ij}^{n} \geq 0;\\
    -(f_{max}^{adh} + k_{adh} \delta_{ij}^{n} ) & \quad -\frac{f_{max}^{adh}}{k_{adh}} \leq \delta_{ij}^{n} < 0;\\
    0 & \quad \text{else},
  \end{cases}
\end{equation}
where $k_{adh}$ is the adhesion ``stiffness" during unloading.
The maximum adhesion force $f_{max}^{adh}$ is defined identical to the pull-off force of the JKR representation of van der Waals interaction \cite{cohesion}: $f_{max}^{adh} = \frac{3}{2} \pi \gamma R_{\mathrm{eff}}$, where $\gamma$ is the surface energy and $R_{\mathrm{eff}} = \frac{R_i R_j}{R_i + R_j}$ is the effective radius of two particles $i$ and $j$ in contact or close proximity. 

The tangential forces (sliding and rolling) are modelled using linear elastic and dissipative forces, where the rolling force is a virtual force, used to calculate the rolling torque. Both the tangential sliding force $f^s$ and rolling torque $M^r$ are assumed to have a yield criterion, truncating the magnitude of $\delta^s$ and $\delta^r$ (the sliding and rolling displacements, respectively) as necessary to satisfy:
$|f^s| \leq \mu_s |f_{ij}^{n}-f_{ij}^{adh}|$ and $\mathbf{M}^r = R_{\mathrm{eff}} \mathbf{n} \times f^r$ with $|f^r| \leq \mu_r |f_{ij}^{n}-f_{ij}^{adh}|$, where $\mu_\mathrm{s}$ and $\mu_\mathrm{r}$ are the sliding and rolling friction coefficients, respectively, usually assumed to be constant (Coulomb type).
More details about the contact model can be found in \cite{luding2008,thomasclouser,thomasfriction}.
\subsection{DPM parameters}\label{sec:modelParameters}
The simulations of the spreading process are done using the open-source code MercuryDPM \cite{mercurydpm}. The same parameter values, see Table\,\ref{table:dpmSimParameters}, were set for particle-particle, particle-substrate and particle-tool interactions. 

The normal spring $k_n$ and damping $\eta_n$ constants are set such that the collision time $t_{\mathrm{c}} = t_{\mathrm{g}} / 200$ with $t{\mathrm{g}} = \sqrt{D_{\mathrm{50}}/g}$ and an intermediate restitution coefficient of $\epsilon = 0.4$ (for no adhesion) is assumed.
To study the effect of particle surface roughness and sphericity, we simulate the spreading process for varying values of interparticle sliding friction $\mu_\mathrm{s}$ and rolling friction $\mu_\mathrm{r}$, as illustrated in section \ref{sec:DOS}.
To study the effect of particle cohesion, we simulate the spreading process for varying values of $\gamma$ such that we can observe a wide range of particle bond numbers.
E.g., for $\gamma$ = 0.1\,$\mathrm{mJ/m^2}$: $Bo_i = \frac{f_{max}^{adh}}{m_i g} = \frac{9 \gamma}{4 \rho_p D_{i}^{2} g} \approx 36, 4, 0.8$, for $D_{min} = 12$, $D_{50} = 37$, and $D_{max} = 79$\,$\mathrm{\mu m}$, respectively; This is consistent with the observation that the effect of cohesion is only moderate in the data presented in section \ref{sec:results} for $\gamma$ = 0.1\,$\mathrm{mJ/m^2}$ ($Bo_{\mathrm{50}}$ = 4). 
The model parameters are set to assure that the interaction is computationally stiff enough  i.e. the particle overlap is well below 1\,\% of particle diameter, preventing unrealistic bulk behaviour.
Table\,\ref{table:dpmSimParameters} shows the main DPM simulation parameters.

A log-normal particle size distribution (PSD) is used, fitted to the particle size distribution of Ti-6Al4V. The PSD is measured using laser diffraction with $D_{10}$ = 24\,$\mathrm{\mu m}$, $D_{50}$ = 37\,$\mathrm{\mu m}$ and $D_{90}$ = 56\,$\mathrm{\mu m}$. Fig.\,\ref{fig:PSD} shows the PSD of Ti-6Al4V as implemented in simulation.

\subsection{Simulation setup}\label{sec:simSetup}
We simulate a small part of the powder bed (width 1\,mm), using periodic boundary condition in $y$-direction. The spreading tool is a blade (commonly used in AM machines) as shown in Fig.\,\ref{fig:simsetup}, moving from left to right at a constant speed $v_\mathrm{T}$. The substrate is assumed to be smooth. We insert particles in front of the spreader tool, at $(x,y,z)$ $\in$ [0.5,2.5]\,mm $\times$ [0,1]\,mm $\times$ [0,$h$]\,mm until the total particle volume equals 0.7 $\mathrm{mm^3}$, which is sufficient material to create a powder layer of 7\,mm length, 1\,mm width and 0.1\,mm height.
After the particles are settled down and the system is relaxed, the simulation of the spreading process starts by moving the tool at a constant speed $v_\mathrm{T}$. Particles reaching the end of the powder bed (at $x$ = 10\,mm) get deleted. The simulation ends after spreading the particles in a layer where the tool gap is always set to $H$ = 100\,$\mathrm{\mu m}$, which corresponds to about 2.7$\times D_{\mathrm{50}}$ in $z$-direction. We stop the simulation at time $t_{\mathrm{max}}$ when the system is static i.e. the kinetic energy is very low.
Fig.\ref{fig:simsetup} shows the simulation setup, after inserted particles have settled down and during the spreading process.

\subsection{Design of simulations} \label{sec:DOS}
The aim of this study is to find the effect of material properties and process parameters on the spread powder layer quality. The parameter values considered for the spreading process simulations are shown in Table\,\ref{table:DOS}.

The upper limit of the interparticle sliding friction $\mu_\mathrm{s}$ is chosen to be 0.5, which is realistic for metal powders \cite{Geer2018,CLEARY2010}.
The rolling friction was varied as a simple measure to mimic particles non-sphericity \cite{Wensrich2012,Wensrich2014} and was varied between 0.005-0.4. 

The process parameters considered in this study are the spreading tool speed and the spreading tool geometry. The basic spreading tool velocity is chosen to be $v_\mathrm{T}$ = 10\,mm/s, which is low enough such that there are no inertia effects on the bulk behaviour of particles during spreading \cite{Meier2019b}. Two different spreading tools are considered, a blade Fig.\,\ref{fig:simsetup} and a counter-clockwise (cc) rotating roller, Fig.\,\ref{fig:simsetupRoller} similar to the blade setup presented previously in section \ref{sec:simSetup}. The roller radius is $r_{roller}$ = 0.5\,mm and the angular velocity is $w_\mathrm{roller} = - v_\mathrm{T} / r_{\mathrm{roller}}$. The gap height is fixed at $H$ = 100\,$\mathrm{\mu m}$, such that it is higher than the maximum particle diameter $D_\mathrm{max}$ = 79\,$\mathrm{\mu m}$ \cite{Mindt2016}. The variation of gap height is not considered here, where increasing the gap height will increase the spread layer packing height and fraction \cite{Nan2018a,Nan2018b,Meier2019b}.

We use a full factorial design simulating the effect of four variables for two spreading tools geometry, Table\,\ref{table:DOS}.
%

\subsection{Powder layer characterisation}\label{sec:layerMetrics}
In sections \ref{sec:MF} and \ref{sec:CG}, we define two different measures to quantify the powder layer characteristics, namely, mass fraction and uniformity, at the end of the simulation, i.e. after spreading.
\subsubsection{Powder layer mass fraction $M\!F$}\label{sec:MF}
To evaluate the powder layer quality, we define the spread layer mass fraction for an assumed volume fraction. Assuming that the layer volume under consideration is $V_{layer} = 7 \times 1 \times 0.1\,\mathrm{mm^3}$, and assuming a theoretical volume fraction $V\!F = 100\,\%$, we can calculate the particle maximum mass needed to achieve $V\!F$,
\begin{equation}
m_{layer} = V\!F \times V_\mathrm{layer} \times \rho_p,
\end{equation}
where $\rho_p$ is the particle density and $m_\mathrm{layer} = 7 \times 1 \times 0.1 \times10^{-6} \times 4430 = 0.003101$\,kg. Then we can define the spread layer mass fraction $M\!F$ as
\begin{equation}
M\!F = \frac{m_\mathrm{SL}}{m_\mathrm{layer}},
\end{equation}
where $m_\mathrm{SL}$ is the total mass of remaining particles after the spreading process within the considered layer volume $V_{layer}$. It should be noted that the maximum volume fraction that can be achieved is about 64\% for random close packing.
%
\subsubsection{Powder layer uniformity and porosity}\label{sec:CG}
To characterise the spread powder layer, continuum fields, e.g., solid volume fraction can be extracted from discrete data using micro-macro transition methods, such as coarse-graining (CG) \cite{CG}. 
This method has the advantage that the fields produced satisfy mass and momentum balance exactly even near the boundaries.
Here we only use the macroscopic solid volume fraction $\phi$ after spreading,
\begin{equation}
\phi(x,y,z) = \sum_{i=1}^{N} V_i \psi (\mathbf{r}-\mathbf{r}_i(t_\mathrm{max})),
\end{equation}
where $V_i$ is particle volume. 
Here, we use a Gaussian coarse-graining function $\psi$ of width (standard deviation) $w = 40\,\mathrm{\mu m}$ and a cut-off $w_c = 3w$. The width was chosen to be approximately the maximum particle radius \cite{Deepack}.
More details of the CG method are beyond the scope of this paper and the interested reader is referred to \cite{CG,Goldhirsch, Deepack}.
Height integration in $z$-direction yields a spatial distribution field of depth-averaged powder layer solid volume fraction in $xy$-directions, 
\begin{equation}
\bar{\phi}(x,y) = \frac{1}{H} \int_{0}^{H} \phi dz,
\end{equation}
where $H$ is the gap height (here also the expected, optimal layer thickness). 
The spatial distribution of the depth-averaged solid volume fraction $\bar{\phi}(x,y)$ can be used as a quantitative measure of the powder layer quality and uniformity, as illustrated in the CG figures in section \ref{sec:resultsLayerUniformity}.
However, we need scalar values for a comprehensive comparison. We define the coefficient of variation ($cv$) for each $\bar{\phi}(x,y)$ distribution, we obtain a scalar value that can be defined as a measure of spread powder layer uniformity. The coefficient of variation is defined as the ratio of the standard deviation $\sigma$ to the mean $\mu_{\bar\phi}$ of $\bar{\phi}(x,y)$:
\begin{equation}
    cv = \frac{\sigma}{\mu_{\bar\phi}},
\end{equation}
where the mean value $\mu_{\bar\phi}$ and standard deviation $\sigma$ of the solid volume fraction $\bar{\phi}(x,y)$ are defined as
\begin{equation}
    \mu_{\bar\phi} = \frac{1}{k} \sum_{i = 1}^{k} \bar{\phi}(x,y),
\end{equation}
\begin{equation}
    \sigma = \frac{1}{k-1} \sum_{i = 1}^{k} |\bar{\phi}(x,y) - \mu_{\bar\phi}|^2,
\end{equation}
where $k = 100 \times 100$ is the number of sampled (square) grid points in $x$ and $y$ directions, respectively. Non-uniform layers have a high $cv$, while relatively uniform ones have a low $cv$. This will be illustrated in section \ref{sec:resultsLayerUniformity}. Fig.\,\ref{fig:CGExample} shows an example of the spatial and probability distributions of the solid volume fraction $\bar{\phi}$.

%% file: results.tex
First, we discuss powder layer defects that reduce powder layer quality in section \ref{sec:defects}. Then, we present and discuss the measured powder layer mass fractions $M\!F$ and uniformity in section \ref{sec:resultsMF} and section \ref{sec:resultsLayerUniformity}, respectively. Finally, we illustrate particle size segregation in section \ref{sec:particlesSegregation}.
\subsection{Spread powder layer defects}\label{sec:defects}
We observe several different powder layer defects that affect the powder layer quality, i.e., increase layer porosity. Those defects include empty patches and vacancies, which are caused by particle drag. This occurs when particles are forced forward by the spreading tool, keeping other particles from flowing through the gap. This can either be due to interlocking/clogging for highly frictional (rough) particles, or agglomeration/sticking for strongly cohesive particles. Fig.\,\ref{fig:defectsBlade} and Fig.\,\ref{fig:defectsRoller} show that particle interlocking, particle drag and particle agglomerates occur for both the blade and the counter-clockwise (cc) rotating roller. Whereas strongly cohesive particles stick on both tools, Fig.\,\ref{fig:defectsRoller}c shows that for the cc rotating roller.

Particle drag during the spreading process was also reported experimentally by Foster et al. \cite{Foster2015} and Abdelrahman et al. \cite{Abdelrahman2017}.

\subsection{Powder layer mass fraction ($M\!F$)}\label{sec:resultsMF}
Next, we try to understand the collective effect of three different material properties on the layer quality, as quantified by the spread layer mass fraction $M\!F$, for different spreading speeds, using the blade in Fig.\,\ref{fig:bladeMF} or the cc rotating roller in Fig.\,\ref{fig:rollerMF}. The major findings and observations are summarized next, while more details are discussed in the following subsections.

The spread layer mass fraction is displayed in Figs.\ref{fig:bladeMF} and Fig.\,\ref{fig:rollerMF} for the blade and cc rotating roller, respectively. Each subplot shows the dependence of $M\!F$ on the rolling and sliding friction, accumulating the results of 42 simulations. The plots are arranged in a 3x3 matrix,  where each row shows a different speed, $v_\mathrm{T} = 10$, 50, 100\,mm/s. and each column shows a different mean cohesiveness, $Bo_{\mathrm{50}} = 0$, 4, 15.
In general, orange indicated very good (close to optimal) layer quality, yellow moderate, deteriorating via green and light to dark blue, which indicates very bad layer quality (almost empty layers).
Snapshots of several representative examples (indicated by red dot markers in Figs.\ref{fig:bladeMF} and Fig.\,\ref{fig:rollerMF}) are displayed in Fig.\,\ref{fig:topViewMFblade} (blade) 
and Fig.\,\ref{fig:topViewMFroller} (cc rotating roller).

From this representation -- just by looking at the dominant color in a subplot -- we can identify parameter combinations that lead to good or bad results in layer quality.
Generally, for either tool, we see that as interparticle friction $\mu_\mathrm{s}$ and $\mu_\mathrm{r}$ increase, $M\!F$ decreases. 
For large values of $\mu_\mathrm{s}$, an increase in rolling friction, $\mu_\mathrm{r}$, decreases $M\!F$, whereas for small values of $\mu_\mathrm{s}$, $\mu_\mathrm{r}$ has only little influence.
In other words, $M\!F$ decreases as particle roughness and non-sphericity increase. 
In addition, increasing spreading speed typically decreases layer quality and $M\!F$. Likewise, increasing particle cohesiveness, $Bo_{\mathrm{50}}$, for constant spreading speed, reduces $M\!F$ -- with some exceptions, as discussed in detail in section \ref{sec:resultsMFBladeSpreader} for the blade and in section \ref{sec:resultsMFRollerSpreader} for the cc rotating roller.
Comparing the two spreader geometries, unlike the blade tool, the cc rotating roller compacts the powder during the spreading process. This results always in a higher spread layer mass fraction, with $M\!F_{max}$ of about 60 \%, while for the blade $M\!F_{max}$ is only about 50 \%. 

For some cases, e.g., Fig.\,\ref{fig:bladeMF}a and Fig.\,\ref{fig:rollerMF}a, we see that at low $\mu_\mathrm{s}$, as $\mu_\mathrm{r}$ increases, $M\!F$ is almost unaffected. In contrast, at low $\mu_\mathrm{r}$, as $\mu_\mathrm{s}$ increases, $M\!F$ reduces considerably. 
This can be related to particle segregation, as will be discussed in detail in section \ref{sec:particlesSegregation}.

\subsubsection{Blade spreader}\label{sec:resultsMFBladeSpreader}
Zooming into the $M\!F$ results using the blade tool, in Fig.\,\ref{fig:bladeMF}, most parameter combinations lead to bad layer quality (small $M\!F$). Only in Fig.\,\ref{fig:bladeMF}a,b yellow/green dominate on bottom and left, whereas in all other subplots $M\!F$ values indicate bad layer quality. 
This means that the layer quality is better, the smaller the sliding friction and it improves a little for smaller rolling friction. 
From Fig.\,\ref{fig:bladeMF}a,b,c we see that increasing particle cohesiveness from $Bo_{\mathrm{50}}$ = 0 to 4 barely affects $M\!F$, while it is deteriorating for strong cohesion, $Bo_{\mathrm{50}}$ = 15.
On the other hand, we obtain relatively high $M\!F$ for non- and weakly cohesive particles ($Bo_{\mathrm{50}} = 0, 4$, respectively) in two cases: (i) when particle roughness is relatively low (low $\mu_\mathrm{s}$), even for low particle sphericity (high $\mu_\mathrm{r}$), (ii) when the particle sphericity is high (low $\mu_\mathrm{r}$) even for high particle roughness (high $\mu_\mathrm{s}$).

For strongly cohesive particles Fig.\,\ref{fig:bladeMF}c, the effect of interparticle friction on $M\!F$ is different. We see that $\mu_\mathrm{r}$ has a major negative influence on $M\!F$. However, as $\mu_\mathrm{s}$ increases, $M\!F$ increases. The reason can be due to the fact that as $\mu_\mathrm{s}$ increases, contacts between particles decreases reducing the effective cohesion. 
Fig.\,\ref{fig:cohesion4BladeEffectMuS} show a top view of the spread powder layer for strongly cohesive particles $Bo_{\mathrm{50}} = 15$. It illustrates that powder layer quality decreases as $\mu_\mathrm{s}$ decreases due to particle agglomerates increase.

Fig.\,\ref{fig:bladeMF}d,e,f show $M\!F$ for non-, weakly and strongly cohesive particles, $Bo_{\mathrm{50}} = 0$, 4, 15, respectively, with spreading speed $v_\mathrm{T}$ = 50\,mm/s.
Similarly, Fig.\,\ref{fig:bladeMF}g,h,i with spreading speed $v_\mathrm{T}$ = 100\,mm/s.
We clearly see that, for non- or weakly cohesive particles, increasing the spreading speed $v_\mathrm{T}$ for all cases has reduced the $M\!F$ compared to the lower spreading speed $v_\mathrm{T}$ = 10\,mm/s.
$M\!F_{max}$ is between 30-20\% at $v_\mathrm{T}$ = 50\,mm/s and $M\!F_{max}$ = 20 \% at $v_\mathrm{T}$ = 100\,mm/s, compared to $M\!F_{\mathrm{max}} \approx$ 50\% at $v_\mathrm{T}$ = 10\,mm/s. Qualitatively, this means increasing spreading speed will reduce layer packing fraction.

For strongly cohesive particles, increasing the spreading speed has the opposite effect. $M\!F$ slightly increased at high $\mu_\mathrm{r}$ and low $\mu_\mathrm{s}$ compared to the lower spreading speed $v_\mathrm{T}$ = 10\,mm/s, as seen from the contour lines. Fig.\,\ref{fig:cohesion4BladeEffectSpeed} show the same effect in a top view of spread powder layer with different spreading speeds $v_\mathrm{T}$ = 10, 50, 100\,mm/s for strongly cohesive particles. 
A possible explanation is that the high shear rate resulting from higher spreading speed broke the interlocking in front of the spreading tool caused by particle agglomerates and high $\mu_\mathrm{r}$, allowing particles to flow. 

\subsubsection{Roller spreader}\label{sec:resultsMFRollerSpreader}
Simulations with the cc rotating roller (Fig.\,\ref{fig:rollerMF}) generally show a higher $M\!F$ than simulations with the blade spreader. 
However, the main qualitative dependencies on the material properties remain the same;
similar to the blade spreader. Fig.\,\ref{fig:rollerMF}a,b show that the interparticle friction effect on $M\!F$ is almost the same for non- and weakly cohesive particles. 
For strongly cohesive particles Fig.\,\ref{fig:rollerMF}c, we also see similar behaviour as a blade spreader low sphericity (high $\mu_\mathrm{r}$) has major negative influence on $M\!F$. As for the blade, $M\!F$ increases as $\mu_\mathrm{s}$ increases; however, the effect is much more pronounced: at the highest values of $\mu_\mathrm{s}$ and $\mu_\mathrm{r}$, a higher spread layer mass fraction $M\!F$ is obtained, compared to the case for a blade spreader.

%
Fig.\,\ref{fig:rollerMF}d,e show $M\!F$ for non- and weakly cohesive particles, respectively, at $v_\mathrm{T}$ = 50\,mm/s. Similarly, Fig.\,\ref{fig:rollerMF}g,h at $v_\mathrm{T}$ = 100\,mm/s.
We see lower values of $M\!F$ for almost all cases compared to lower spreading speed at $v_\mathrm{T}$ = 10\,mm/s.
Unlike the case of the blade spreader, we see a higher dependency on $\mu_\mathrm{s}$ than on $\mu_\mathrm{r}$ at $v_\mathrm{T}$ = 50\,mm/s. Beside that at low $\mu_\mathrm{s}$, $M\!F$ is higher for weakly cohesive particles compared to non-cohesive ones; which can be seen at $v_\mathrm{T}$ = 100\,mm/s as well. It seems that the combined effect of particle cohesiveness and roller compaction allowed particles to adhere better to each other and to the substrate, compared to non-cohesive particles.

Fig.\,\ref{fig:rollerMF}f,i show $M\!F$ for strongly cohesive particles at spreading speed $v_\mathrm{T}$ = 50 and 100\,mm/s, respectively. At high $\mu_\mathrm{r}$ and low $\mu_\mathrm{s}$, increasing the spreading speed $v_\mathrm{T}$ has significantly increased $M\!F$ compared to lower $v_\mathrm{T}$ = 10\,mm/s. 
Surprisingly, $M\!F$ increases at the higher limit of interparticle friction $\mu_\mathrm{s}$ and $\mu_\mathrm{r}$ compared to intermediate values. This indicates a lower and an upper limit of particle roughness ($\mu_\mathrm{s}$ value) at which particles sphericity ($\mu_\mathrm{r}$ value) has an effect on the spread powder layer quality when using a roller. Fig.\,\ref{fig:cohesion4RollerEffectSpeed} show top view of spread layer using a roller, illustrating the effect of increasing spreading speed $v_\mathrm{T}$. In addition, there is a waving effect on the spread layer for weakly and strongly cohesive particles at low $\mu_\mathrm{s}$ when $v_\mathrm{T}$ = 100\,mm/s e.g. Fig.\,\ref{fig:cohesion4RollerEffectSpeed}f. 

\subsection{Powder layer uniformity and porosity}\label{sec:resultsLayerUniformity}
Dense, uniform powder layers are required to achieve high quality products with low porosity. Previously, we illustrated layer defects which reduce layer uniformity and increase porosity.
In this section, we study the spatial distribution of the powder layer solid volume fraction, $\bar{\phi}(x,y)$, where it is utilized to evaluate powder layer uniformity.

In Fig.\,\ref{fig:CGmapsBladeVT10NoC}, we show the solid volume fraction $\bar\phi$ for three cases, that are representative of the typical kind of powder layers we obtain after spreading: (i) a uniform layer, (ii) empty patches and (ii) a nearly empty layer. We further plot the probability distribution for each case and determine its coefficient of variation $cv$. 
The cases shown use a blade spreader at $v_\mathrm{T}$ = 10\,mm/s with weakly cohesive particles, $Bo_{\mathrm{50}}$ = 4. We see good powder layers with a homogeneous narrow normal distribution at low interparticle friction; as the interparticle friction increases the spread powder layer uniformity decreases, with empty patches indicated by the dark blue regions in the contour plot of $\bar\phi$. For very high interparticle friction, the probability distribution shows a high peak at zero. 
Predictably, we observe the highest coefficients of variation for nearly empty layers, and the lowest for uniform layers; thus, we aim to use the coefficient of variation as a measure of layer uniformity and will see if and how it correlates with $M\!F$.

Figs.\ref{fig:cvBlade} and \ref{fig:cvRoller} show the correlation between $cv$ and $M\!F$, for a blade and a cc rotating roller, respectively. Each row shows one spreader velocity, increasing from top to bottom. 
The value of $cv_u$ = 0.2 is set as an upper limit for a uniform layers. For increasing $cv>cv_u$ the layer uniformity decreases, where layer vacancies and empty patches occur, up to the upper limit of $cv_m$ = 0.9; for $cv>cv_m$, we see severe particle interlocking and drag causing empty layers. In Fig.\,\ref{fig:cvAll}, we see that the data for different $Bo_{50}$ accumulate on a master curve, sometimes well above the upper, in between, or well below the lower limits. The former case are bad packings, whereas the latter are the good quality layers.

Thus, we propose the following function to fit the data for a blade and cc rotating roller at \! $v_\mathrm{T} = 10$\,mm/s
\begin{equation} \label{eq:fit}
    cv_\mathrm{fit} = \left( \left(\frac{M\!F}{d_1} \right)^{p_1} + \left( \frac{M\!F}{d_2} \right)^{p_2} \right)^{-1} \,.
\end{equation}
First assuming $d_1 = 20$ and $p_1 = 0.5$, which reasonably fits the large $cv>cv_m$ data, we get $p_2 = 3 \pm 0.4$, $d_2 = 20.5 \pm 1$ or $p_2 = 3 \pm 0.2$, $d_2 = 23.5 \pm 1$, for a blade or cc rotating roller, respectively. This master curves fit all these data pretty well and differ only slightly in the $d_2$ coefficient, meaning that the roller produces slightly better layers with higher $M\!F$, or equivalently lower $cv$.

In more detail, for a blade spreader, Fig.\,\ref{fig:cvBlade}, we see that powder layers with high $M\!F$ are uniform, mostly for non- and weakly cohesive particles, $Bo_\mathrm{50} = 0$ and 4, respectively. In contrast, for strongly cohesive particles $Bo_\mathrm{50} = 15$, powder layers either end up close to the low uniformity limit, $cv_u$, or are of very bad quality. Increasing the spreading speed from $v_\mathrm{T} = 10$\,mm/s to $v_\mathrm{T} = 50$ and 100\,mm/s, decreases layer uniformity significantly in almost all cases.

For a cc rotating roller, Fig.\,\ref{fig:cvRoller}, we see similar qualitative results as for a blade spreader when changing the parameters. However, increasing the spreading speed has considerably improved powder layer quality for strongly cohesive particles $Bo_\mathrm{50} = 15$.

\subsection{Particle size segregation}\label{sec:particlesSegregation}
Polydisperse particles segregate when they are in motion. Fig.\,\ref{fig:bladepsd} shows the fraction of particles of a given radius remaining in the power layer after spreading. I.e., a value of 60\% indicates that  40\% of the particles have been dragged off the plate by the spreading tool. The data shown is for a blade spreader at $v_\mathrm{T}$ = 10\,mm/s. In Fig.\,\ref{fig:bladepsdMuR} $\mu_\mathrm{r}$ = 0.005 is fixed and $\mu_\mathrm{s}$ is varied. While in Fig.\,\ref{fig:bladepsdMuS} $\mu_\mathrm{s}$ = 0.1 is fixed and $\mu_\mathrm{r}$ is varied. The columns show different bond numbers, from left to right $Bo_{\mathrm{50}}$ = 0, 4 and 15, respectively.
We can clearly see that, when $\mu_\mathrm{r}$ is fixed and as $\mu_\mathrm{s}$ increases Fig.\,\ref{fig:bladepsdMuR}, the retained fraction of small particles increases and the retained fraction of large ones decreases, indicating particle segregation (large particles are more likely to be dragged away by the spreader). In the reverse case, when $\mu_\mathrm{s}$ is fixed and as $\mu_\mathrm{r}$ increases Fig.\,\ref{fig:bladepsdMuS}, we do not see much difference in the retained particle fraction. Thus, $\mu_\mathrm{s}$ has larger effect on size-segregation during the spreading process than $\mu_\mathrm{r}$. For strongly cohesive particles, the behaviour is similar, but less pronounced, due most likely to the formation of particle agglomerates and a higher dependency of $M\!F$ on $\mu_\mathrm{r}$, as mentioned in the previous sections. In addition, It should be noted that the effect of $\mu_\mathrm{r}$ and $\mu_\mathrm{s}$ on particle size segregation is comparable for powder layers within the same range of $M\!F$.

Fig.\,\ref{fig:bladeSegregation} shows the powder heap during the spreading process with a blade. Only very large ($D$ = 75-79\,$\mathrm{\mu m}$) and very small particles ($D$ = 12-24\,$\mathrm{\mu m}$) are made visible, to be able to see particles migration. Fig.\,\ref{fig:bladeSegregation}a shows the powder heap at low $\mu_{\mathrm{r}}$ and high $\mu_{\mathrm{s}}$ for non-cohesive particles, we see that mainly large particles are accumulated at the tip of the powder heap. While at low $\mu_{\mathrm{s}}$ Fig.\,\ref{fig:bladeSegregation}b, both large and small particles are accumulated at the tip of the powder heap. Similar effect can be seen for strongly cohesive particles Fig.\,\ref{fig:bladeSegregation}c,d, however with less frequency due to particle agglomerates.
Similar behaviour is obtained when using a cc rotating roller as a spreading tool, see Fig.\,\ref{fig:rollerpsd} and Fig.\,\ref{fig:rollerSegregation}.

%% file: conclusions.tex
We have simulated the spreading process in AM with the discrete particle method (DPM) and characterized the powder layer quality.
Low powder layer quality is due to powder layer defects. Those include particle drag, interlock and agglomerates, all of which lead to empty patches that reduce packing fraction, uniformity and thus increase layer porosity. Powder layer defects are more likely to occur due to either high rolling friction, $\mu_\mathrm{r}$, high sliding friction, $\mu_\mathrm{s}$, strong particle cohesion, $Bo_{\mathrm{50}}$, or the combined effects of those three material characteristics.

When using a blade as a spreading tool and for non- and weakly cohesive particles ($Bo_{\mathrm{50}} = 0$ and 4, respectively), we obtain relatively uniform layers with high layer mass fractions at either low $\mu_\mathrm{r}$ (even for high $\mu_\mathrm{s}$) or low $\mu_\mathrm{s}$ (even for high $\mu_\mathrm{r}$). For strongly cohesive particles ($Bo_{\mathrm{50}} = 15$), $\mu_\mathrm{r}$ has a major negative influence on layer mass fraction and uniformity, while $\mu_\mathrm{s}$ has a surprising positive effect: we obtain relatively high layer mass fractions and good uniformity only in the limit of low $\mu_\mathrm{r}$ or high $\mu_\mathrm{s}$. 

When using a counter-clock wise rotating roller as a spreading tool, better packed and more uniform layers were obtained for almost all cases due to a constructive compression/shear effect. Similar to the blade tool, for strongly cohesive particles at low $\mu_\mathrm{r}$ or high $\mu_\mathrm{s}$, we see good quality layers. 

Increasing the spreading speed has reduced the layer quality for non- and weakly cohesive particles for both tools. While, for strongly cohesive particles the layer quality slightly improved for a blade spreader, but significantly improved for a cc rotating roller.

For both tools, it was shown that $\mu_\mathrm{s}$ has more influence than $\mu_\mathrm{r}$ on particle size segregation, which is at the origin of some of the non-intuitive trends we observe.

The present work has provided insight into the combined effects of powder material properties and process parameters on the spreading process in AM. Further research will focus on experimental calibration and validation of the spreading process for the powder during the process cycle (virgin, used, recycled, etc). For calibration, the particle and contact properties should be chosen to match the static and dynamic angle of repose, the cohesive strength, and the apparent density of the powder material -- all tests under low confining stress. For validation, the mass fraction of the spread powder layer should be measured from experiments and compared with simulations. All this can be done for virgin and re-used particles. Finally, a microscope can be used to investigate the powder layer uniformity, in order to obtain a solid volume fraction distribution, and thus the coefficient of variation.

%% file: tables.tex
\newpage
\begin{table}[H]
\center
\begin{tabular}{lllll}
\hline
Variable                           & Symbol      & Unit              & Value & Values range \\ \hline
Particle density                   & $\rho_p$      & $\mathrm{kg/m^3}$  & 4430   & -         \\
Normal stiffness                   & $k_n$       & $\mathrm{kg/s^2}$  &  2.2     & - \\
Normal dissipation                 & $\eta_n$       & $\mathrm{kg/s}$  &  3.3$\times \mathrm{10^{-6}}$   & - \\
Friction stiffness                 & $k_{s}^{t}$ & $\mathrm{kg/s^2}$  &  $(2/7) k_n$     & -  \\
Tangential dissipation             & $\eta_{s}^{t}$ & $\mathrm{kg/s}$  &  $(2/7)\eta_n$     & -   \\
Rolling stiffness                  & $k_{r}^{t}$ & $\mathrm{kg/s^2}$  &  $(2/5) k_n$ & -          \\
Rolling dissipation                & $\eta_{r}^{t}$ & $\mathrm{kg/s}$  &  $(2/5)\eta_n$ & -   \\
Particle diameter    & $D_{\mathrm{P}}$           & $\mathrm{\mu m}$   & -  & 12-79 \\
                     & $D_{\mathrm{10}}$         & $\mathrm{\mu m}$   & 24 & -\\
                     & $D_{\mathrm{50}}$         & $\mathrm{\mu m}$   & 37 & -\\
                     & $D_{\mathrm{90}}$         & $\mathrm{\mu m}$   & 56 & -\\
Surface energy                     & $\gamma$    & $\mathrm{mJ/m^2}$  & -   & 0, 0.1, 0.4       \\
Adhesion stiffness                & $k_{adh}$  & $\mathrm{kg/s^2}$    & - & 0.5$k_n$   \\
Number of particles                & $N$           & -                 & 17169  & -            \\
Gap height                         & $H$           & $\mathrm{\mu m}$           & 100  & -  \\
\hline
\end{tabular}
\caption{DPM simulation parameters.}
\label{table:dpmSimParameters}
\end{table}
%
\begin{table}[H]
\center
\begin{adjustbox}{max width=1\textwidth,center}
\begin{tabular}{lllll}
\hline
Variable & Symbol & Values \\ \hline
Sliding friction coefficient & $\mu_\mathrm{s}$ & 0.5, 0.4, 0.3, 0.25, 0.2, 0.1, 0.05     \\
Rolling friction coefficient & $\mu_\mathrm{r}$ & 0.4, 0.3, 0.2, 0.1, 0.05, 0.005\\
Bond number  & $Bo_{\mathrm{50}}$   & 0, 4, 15       \\
Spreading tool speed (mm/s) & $v_\mathrm{T}$ & 10, 50, 100      \\ 
Spreading tool geometry & - & blade, cc rotating roller \\
\hline
\end{tabular}
\end{adjustbox}
\caption{Design of simulation parameter study.}
\label{table:DOS}
\end{table}

%% file: figures.tex
\newpage
\begin{figure}[H]
\centering
\includegraphics[scale=0.35,angle=0]{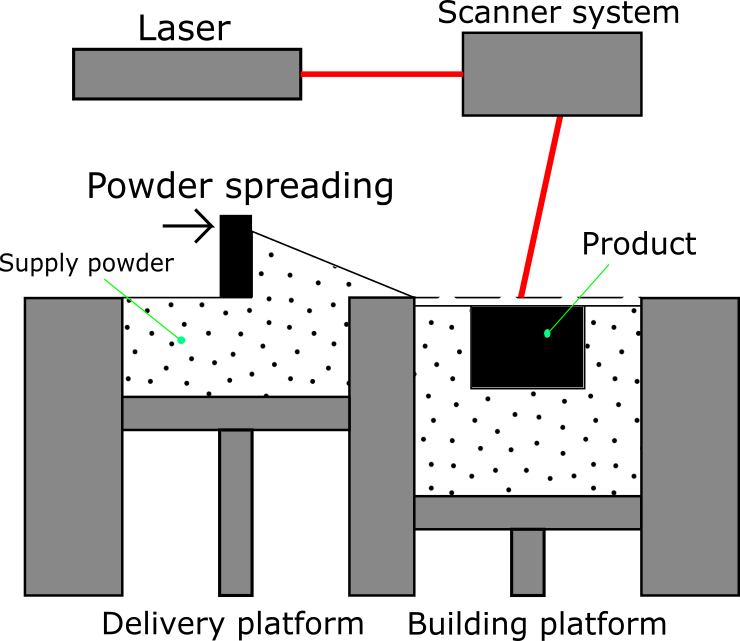}
\caption{Laser powder bed fusion (LPBF) process schematic.}
\label{schematic}
\end{figure}
\begin{figure}[H]
\centering
\begin{subfigure}[t]{0.49\textwidth}
\includegraphics[width=\textwidth]{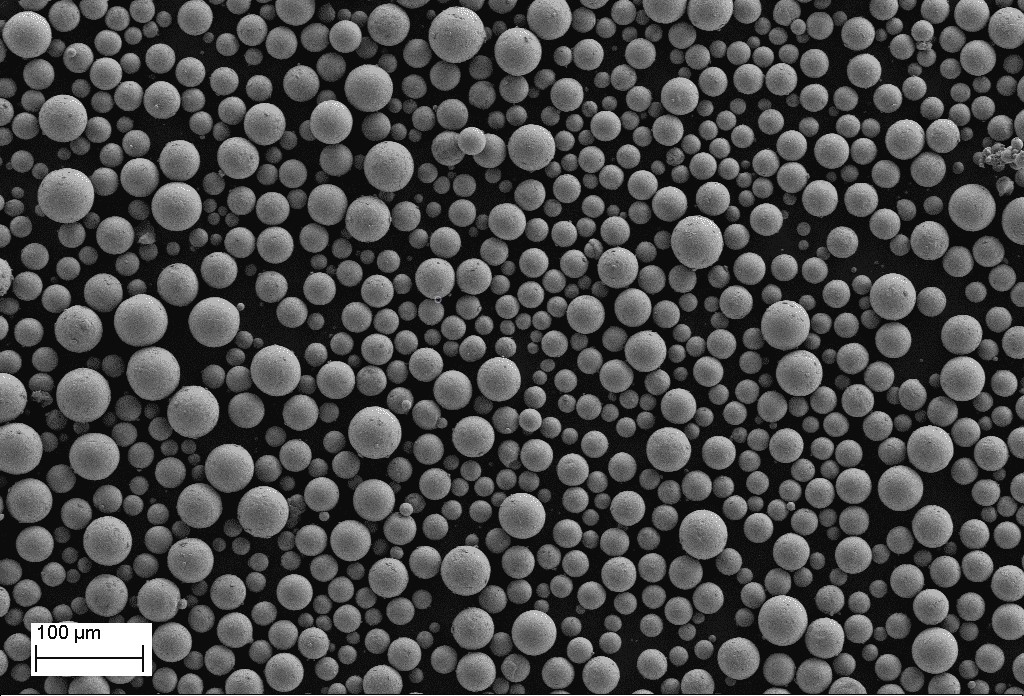}
\caption{Ti-6Al-4V powder produced by plasma rotating electrode. $D_{10}$ = 24 $\mathrm{\mu m}$, $D_{50}$ = 37 $\mathrm{\mu m}$, $D_{90}$ = 56 $\mathrm{\mu m}$.}
\label{fig:Ti15}
\end{subfigure}
\hfill
\begin{subfigure}[t]{0.49\textwidth}
\includegraphics[width=\textwidth]{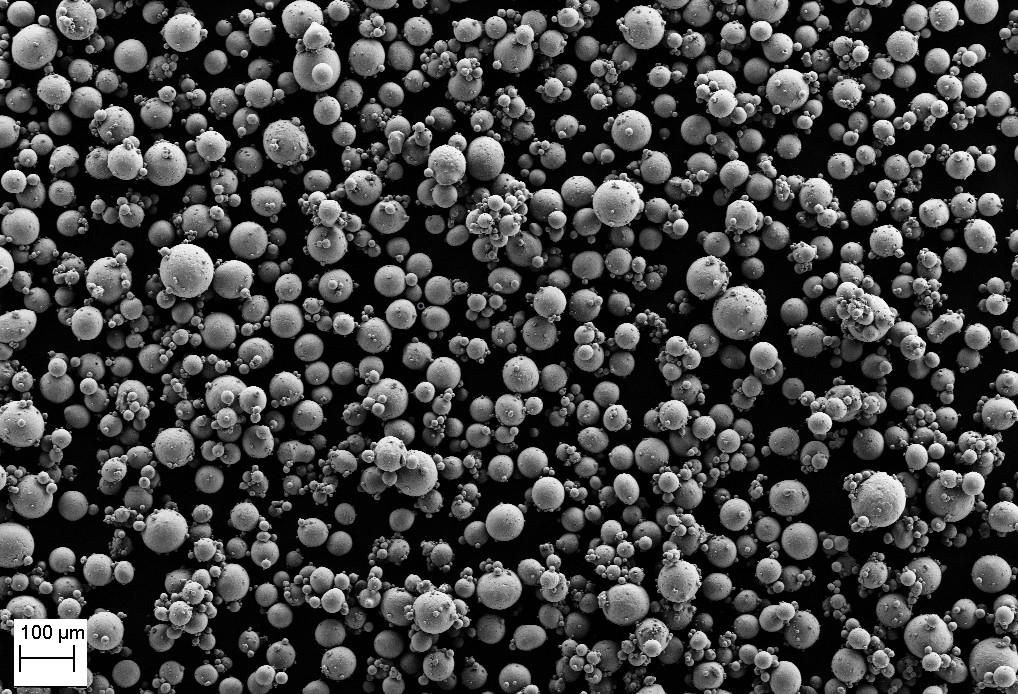}
\caption{Ti-6Al-4V powder produced by gas atomization. $D_{10}$ = 44 $\mathrm{\mu m}$, $D_{50}$ = 70 $\mathrm{\mu m}$, $D_{90}$ = 107 $\mathrm{\mu m}$.}
\label{fig:Ti45}
\end{subfigure}
\caption{SEM image of Ti-6Al-4V powders.}
\end{figure}
\begin{figure}[H]
\centering
\begin{subfigure}[t]{0.32\textwidth}
\includegraphics[width=\textwidth]{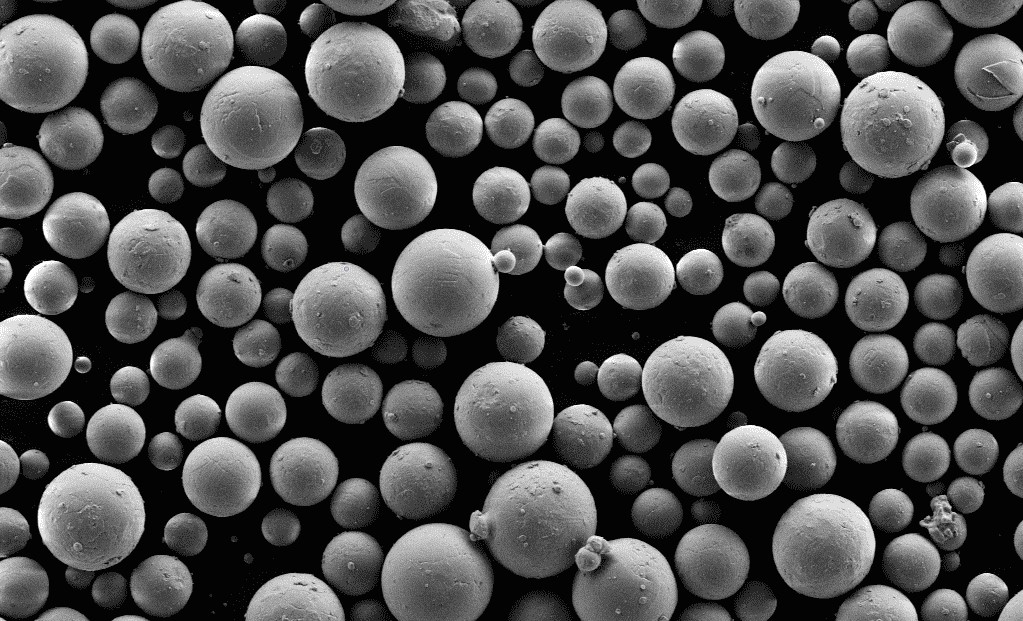}
\caption{Satellites.}
\end{subfigure}
\begin{subfigure}[t]{0.32\textwidth}
\includegraphics[width=\textwidth]{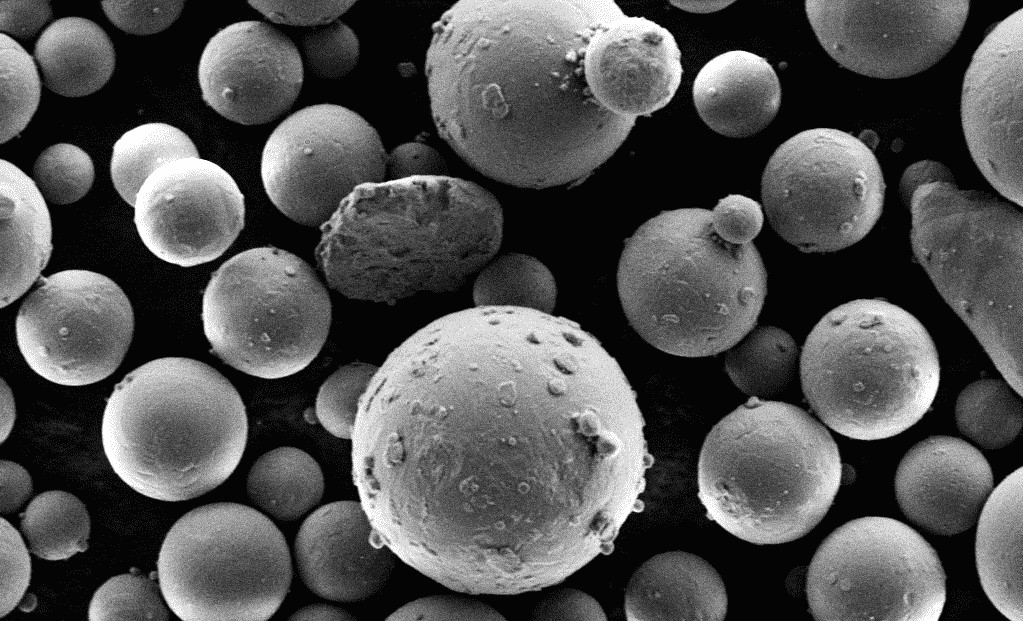}
\caption{Rough surface.}
\end{subfigure}
\begin{subfigure}[t]{0.32\textwidth}
\includegraphics[width=\textwidth]{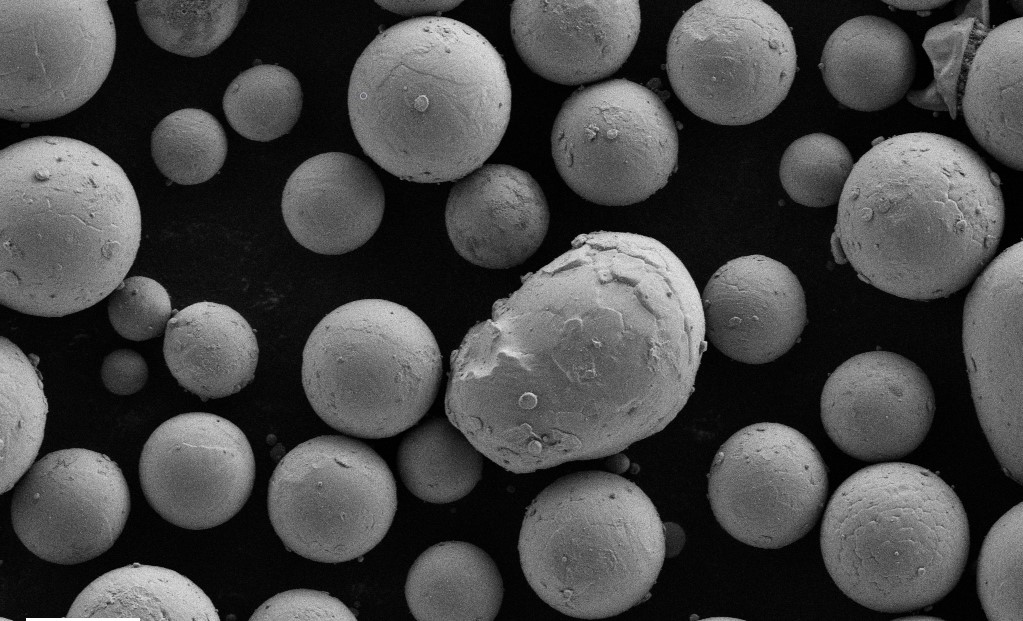}
\caption{Non-sphericity.}
\end{subfigure}
\caption{SEM image of Ti-6Al-4V powder (produced by plasma rotating electrode), showing various types of irregularities.}
\label{fig:Ti15defects}
\end{figure}
%
\begin{figure}[H]
\centering
\begin{subfigure}[t]{0.49\textwidth}
\includegraphics[width=\textwidth]{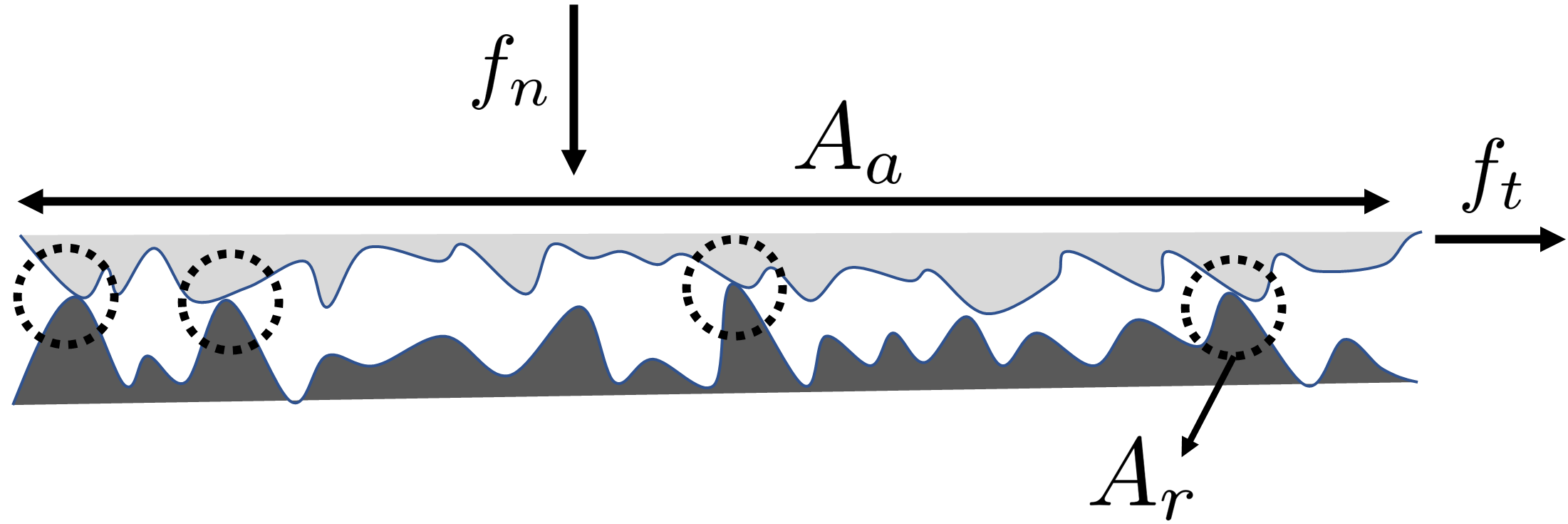}
\caption{Two rough solid surfaces in contact.}
\label{fig:sliding}
\end{subfigure}
\hfill
\begin{subfigure}[t]{0.49\textwidth}
\includegraphics[width=\textwidth]{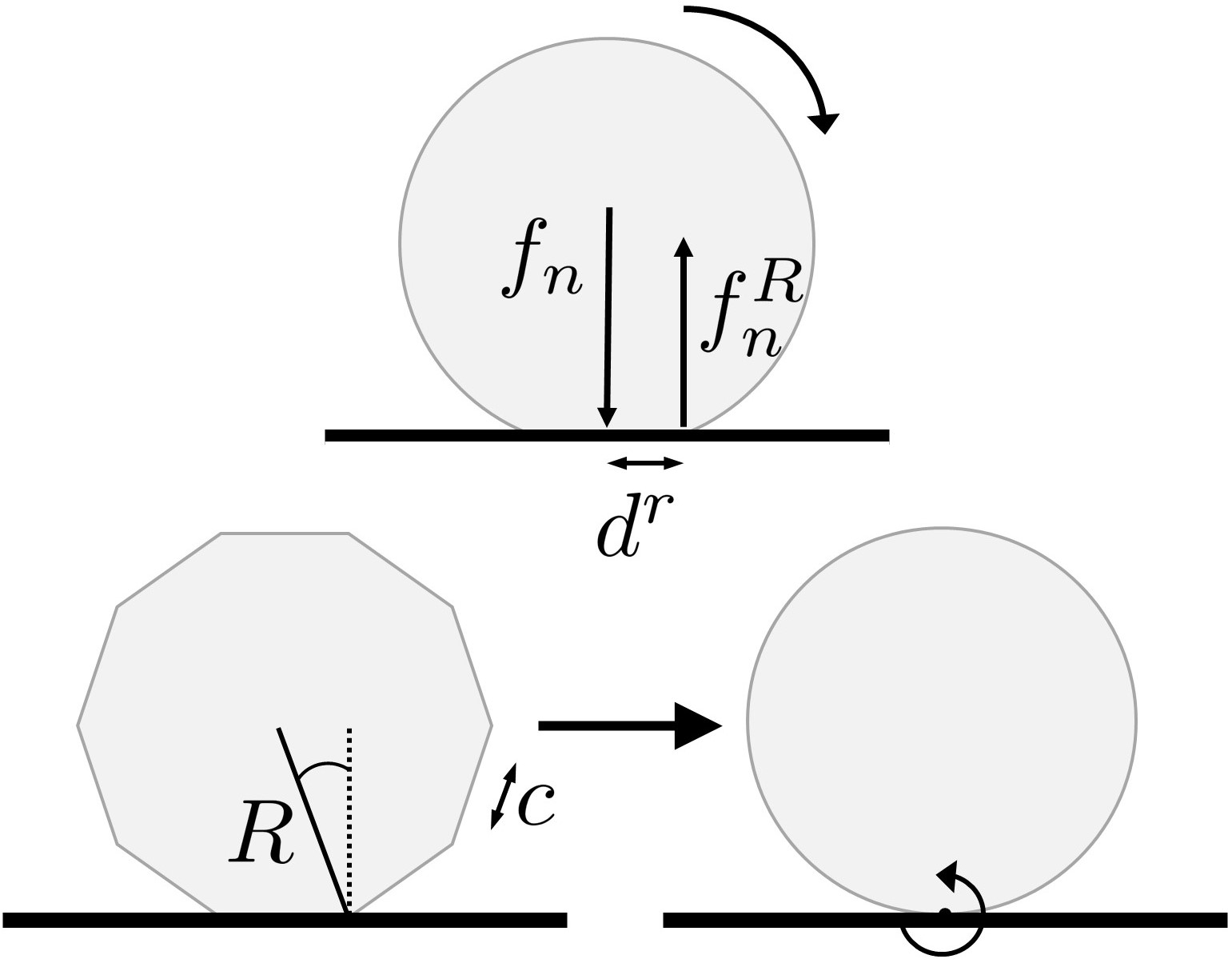}
\caption{The origin of rolling friction and how it can be used to model non-spherical particle shape.}
\label{fig:rolling}
\end{subfigure}
\caption{The physical representation of interparticle sliding and rolling friction.}
\end{figure}
\begin{figure}[H]
\centering
\includegraphics[width=\textwidth]{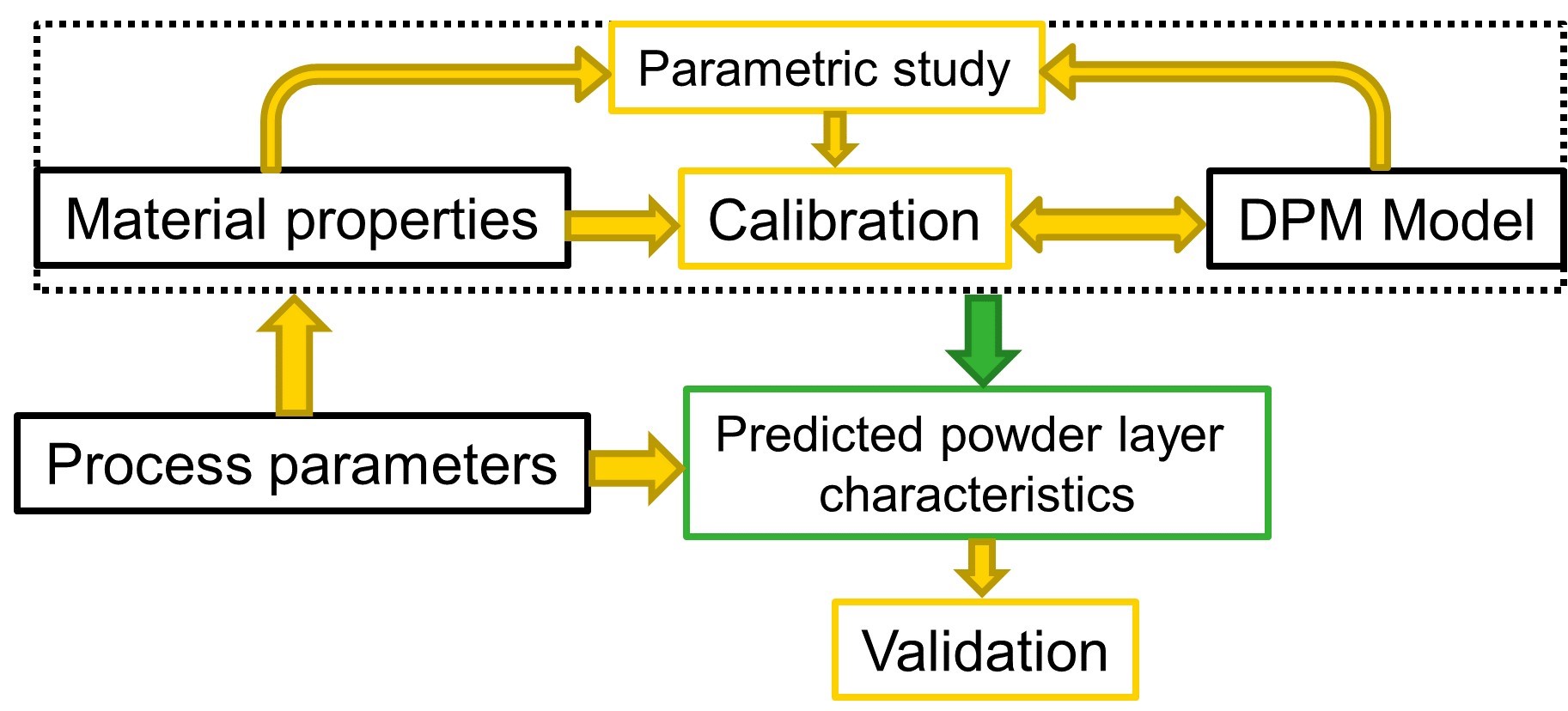}
\caption{DPM simulation, calibration and validation framework.}
\label{fig:simFramework}
\end{figure}
%
\begin{figure}[H]
\centering
\includegraphics[scale=0.5,angle=0]{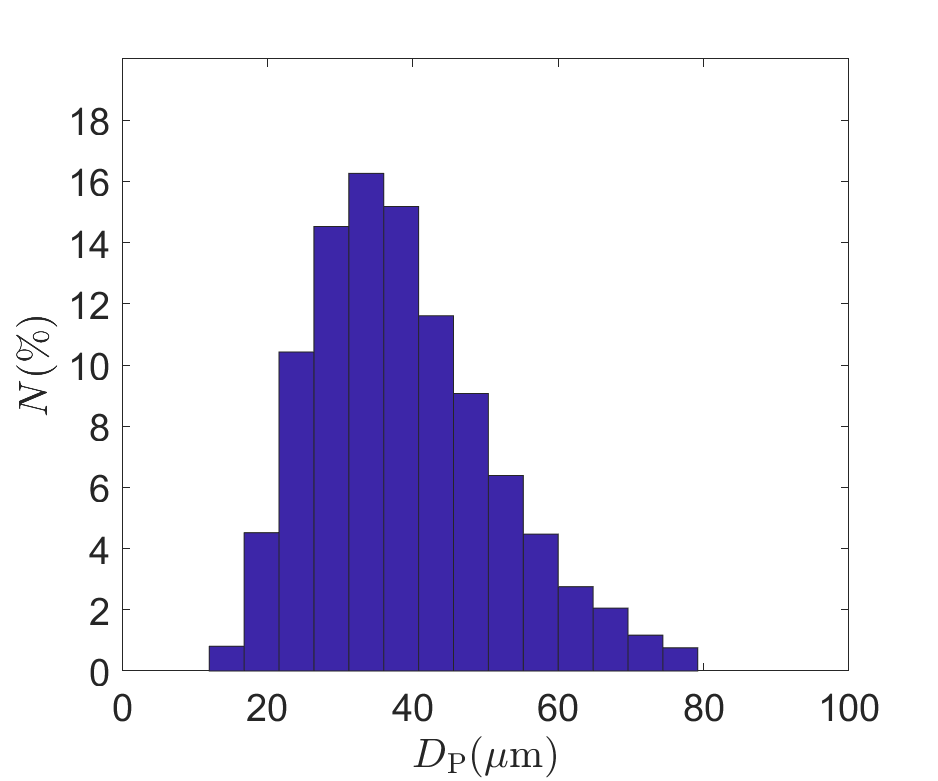}
\caption{Particle size distribution of Ti-6Al-4V as implemented in simulation.}
\label{fig:PSD}
\end{figure}
%
\begin{figure}[H]
\centering
\includegraphics[width=\textwidth]{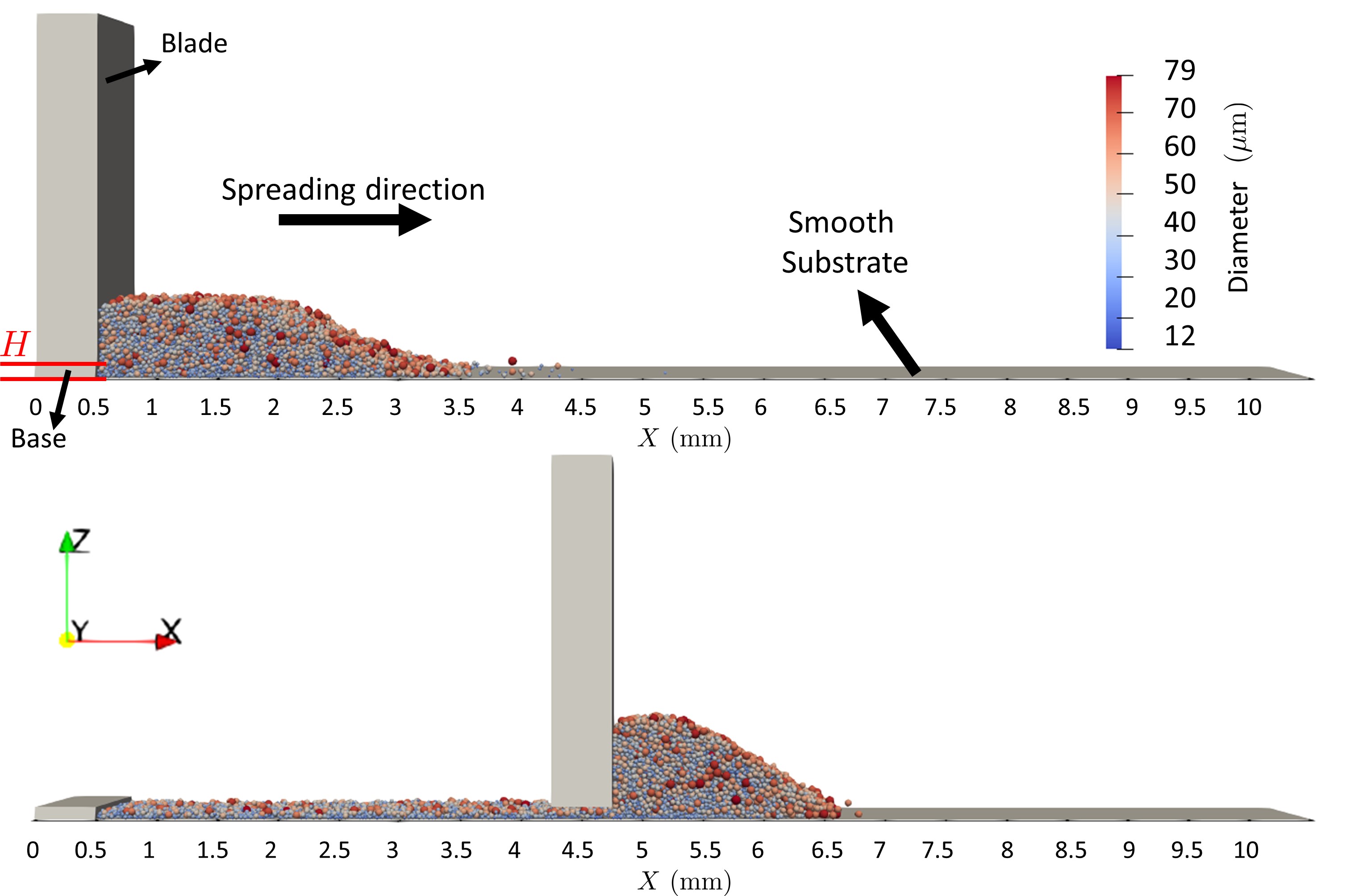}
\caption{Simulation setup using a blade as spreading tool with $Bo_{\mathrm{50}} = 0$, $\mu_\mathrm{r} = 0.1\,\mathrm{\mu m}$, $\mu_\mathrm{s} = 0.5\,\mathrm{\mu m}$ and $v_\mathrm{T}$ = 10\,mm/s.; color indicates particles diameter.}
\label{fig:simsetup}
\end{figure}
\begin{figure}[H]
\centering
\includegraphics[width=\textwidth]{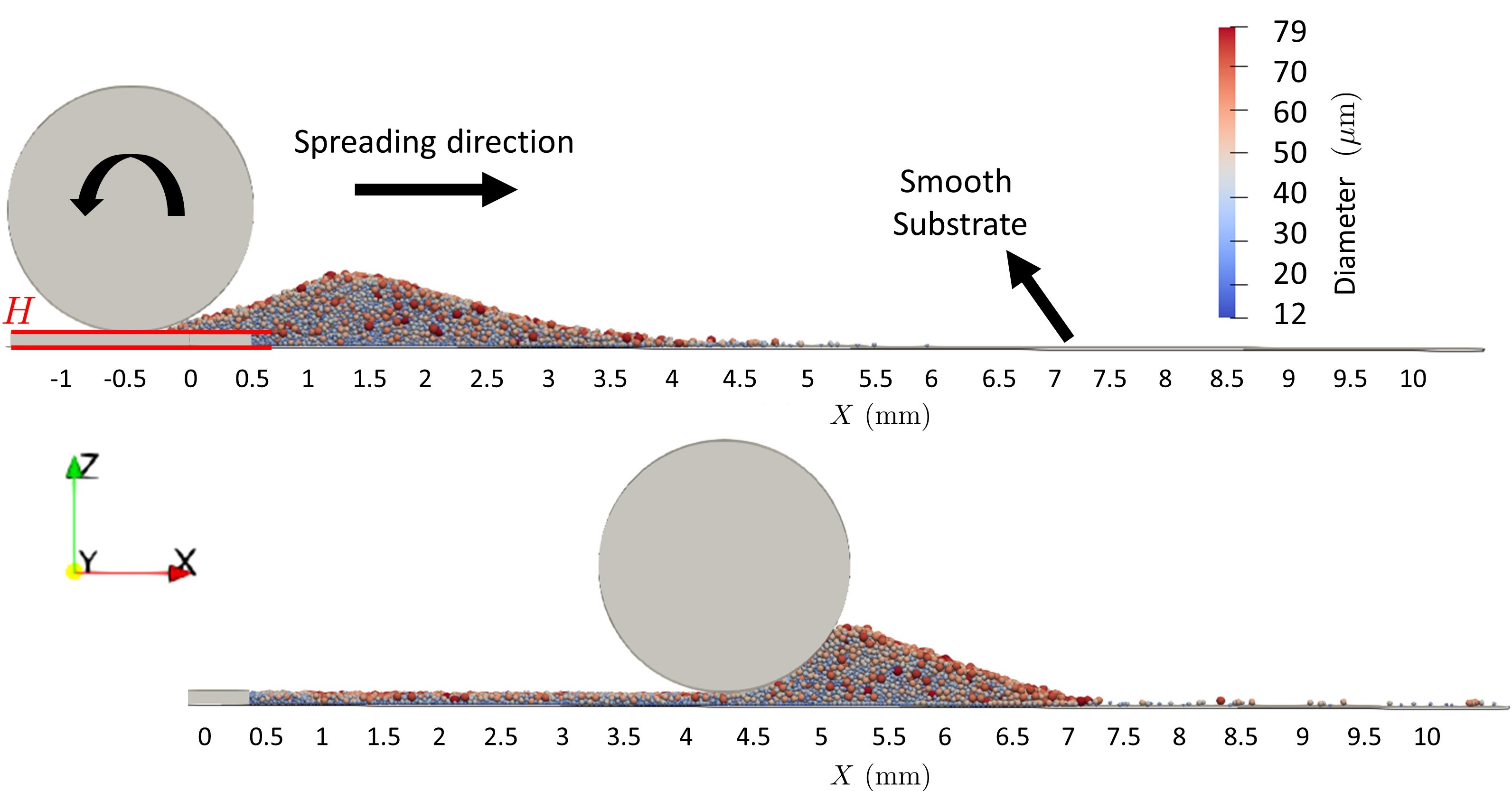}
\caption{Simulation setup using a cc rotating roller as spreading tool with $Bo_{\mathrm{50}} = 0$, $\mu_\mathrm{r} = 0.1\,\mathrm{\mu m}$, $\mu_\mathrm{s} = 0.5\,\mathrm{\mu m}$ and $v_\mathrm{T}$ = 10\,mm/s.}
\label{fig:simsetupRoller}
\end{figure}
\begin{figure}[H]
\centering
\includegraphics[width=\textwidth]{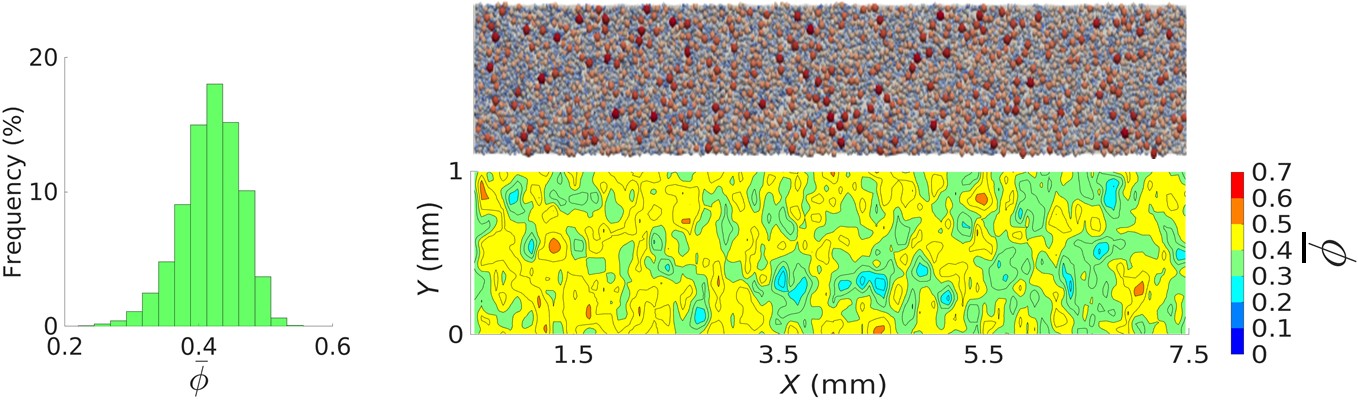}
\caption{Snapshot, spatial distribution and probability distribution of the spread powder layer solid volume fraction $\bar{\phi}$ with a blade with $v_\mathrm{T}$ = 10\,mm/s, for weakly cohesive particles $Bo_{\mathrm{50}}$ = 4 and $\mu_\mathrm{r}$ = 0.005, $\mu_\mathrm{s}$ = 0.1. $\sigma \approx 0.04$, $\mu_{\bar\phi} \approx 0.4$ and $cv \approx 0.1$.}
\label{fig:CGExample}
\end{figure}
%
\begin{figure}[H]
\centering
\includegraphics[width=\textwidth]{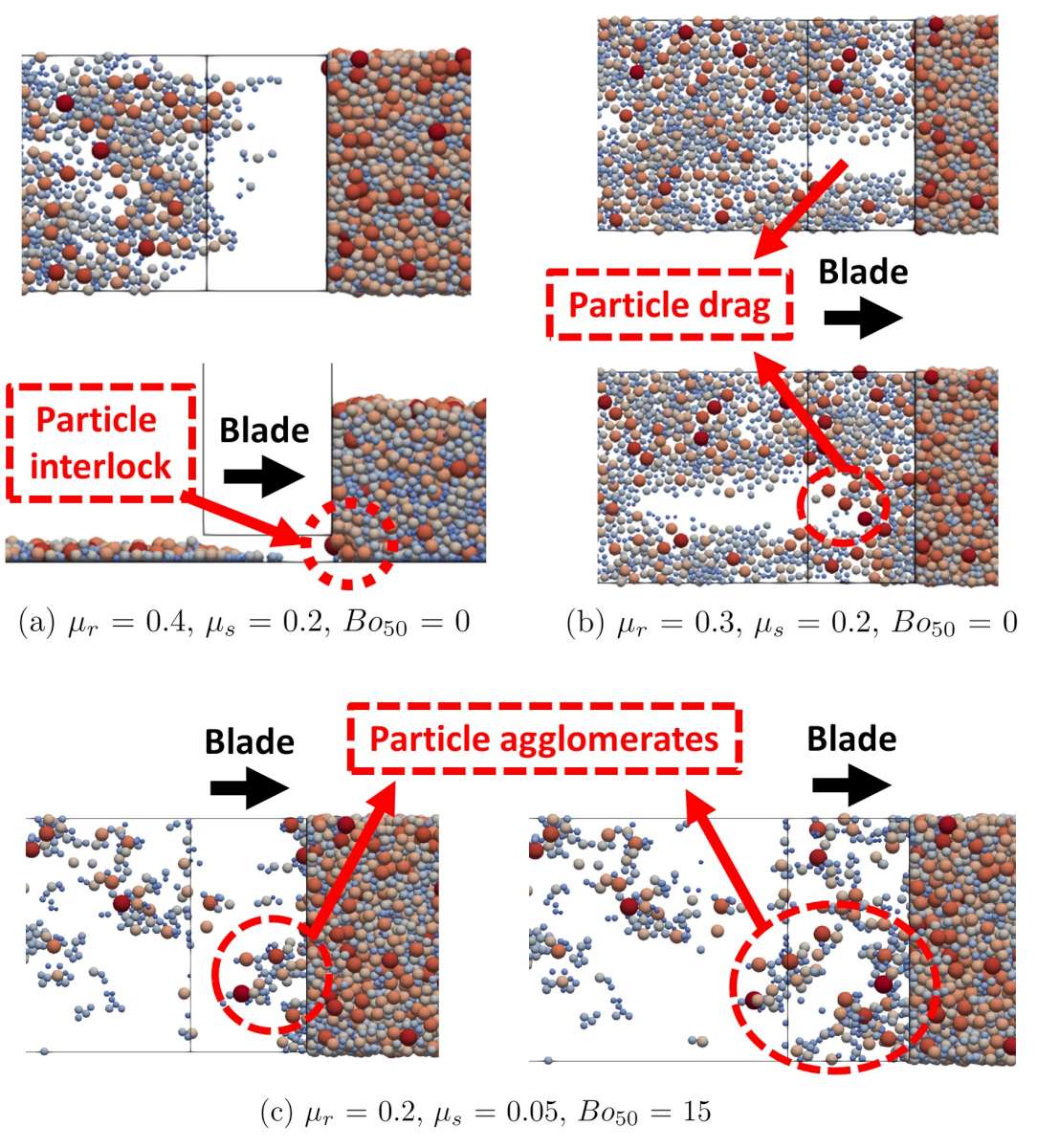}
\caption{Spread powder layer defects for a blade. (a) Particle interlock along the whole layer, top and side views. (b) Particle interlock locally (particle drag), top view at $t=0.29$\,s and $t=0.36$\,s, where the interlock caused by large particles broke. (c) Particle agglomerates for strongly cohesive particles.}
\label{fig:defectsBlade}
\end{figure}
\begin{figure}[H]
\centering
\includegraphics[width=\textwidth]{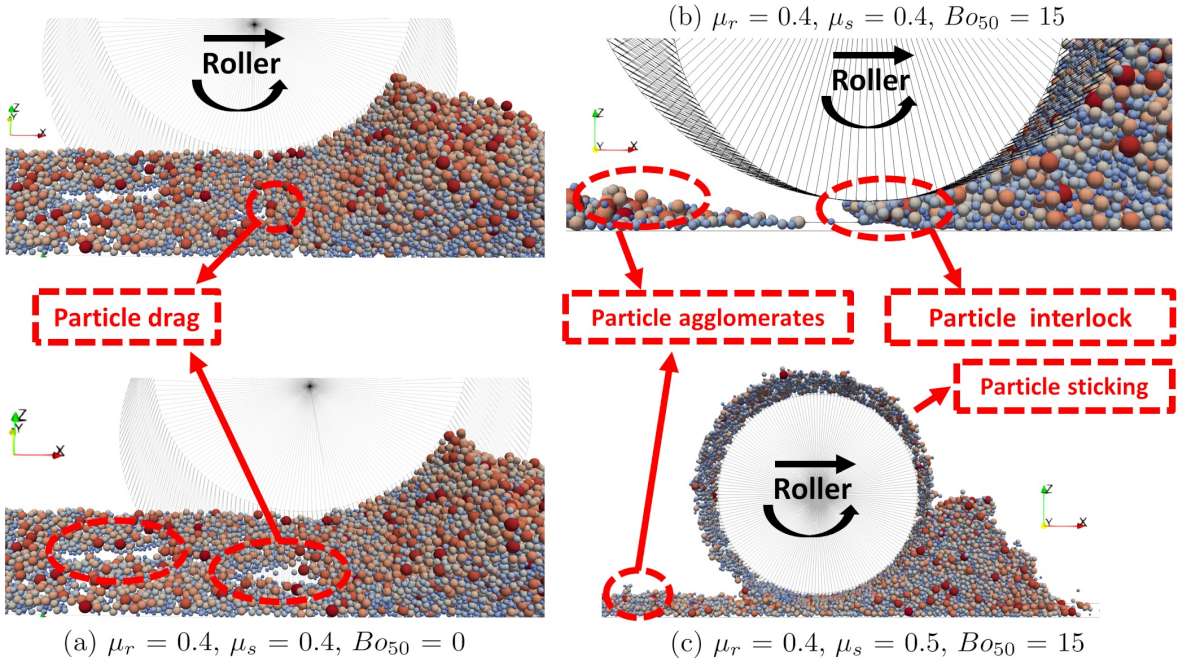}
\caption{Spread powder layer defects for a cc rotating roller. (a) Particle drag at two places, side views. (b) Particle interlock and agglomerates. (c) Particle sticking for strongly cohesive particles.}
\label{fig:defectsRoller}
\end{figure}
%
\begin{figure}[H]
\centering
\includegraphics[width=\textwidth]{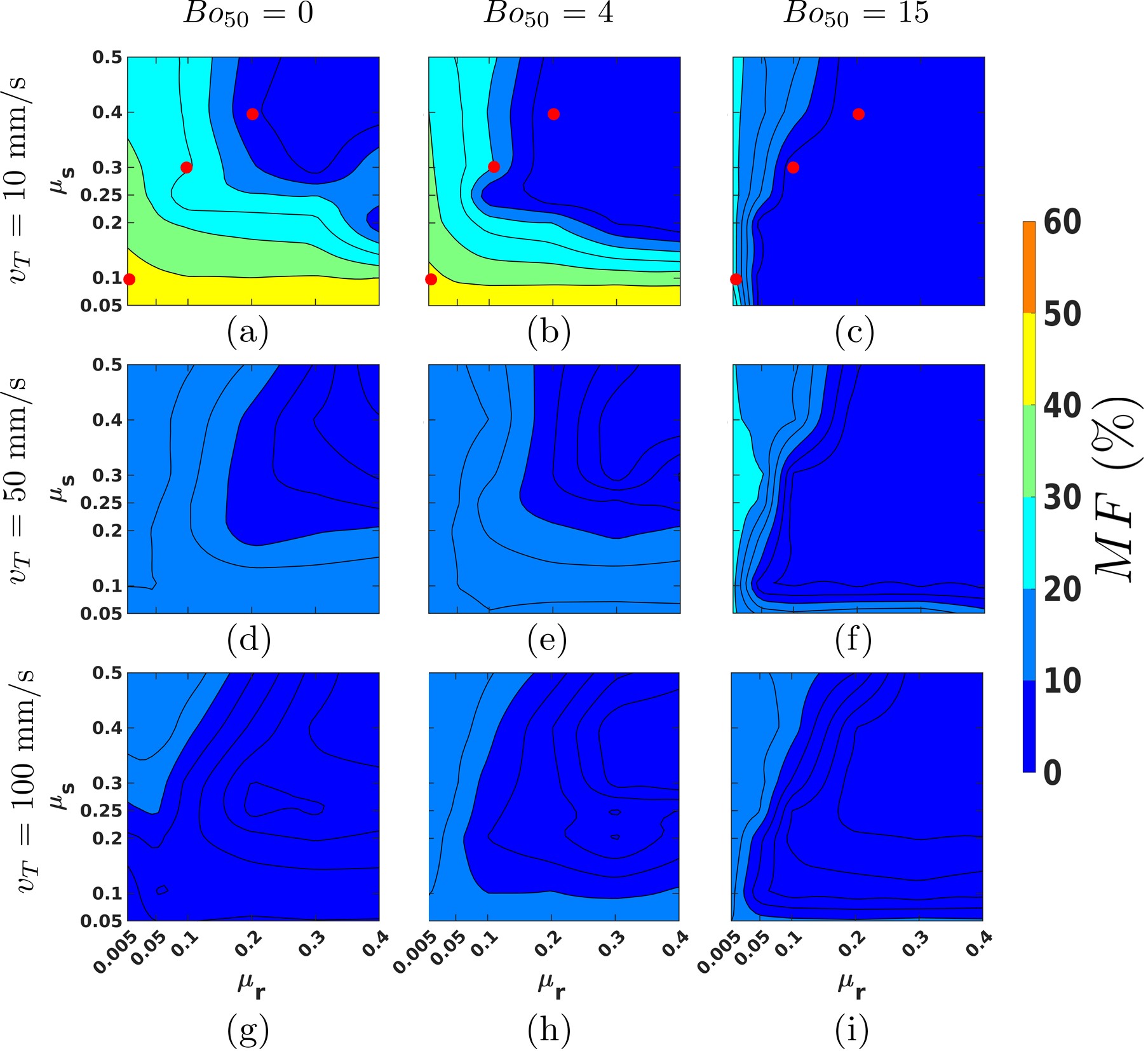}
\caption{Spread layer mass fraction $M\!F$ using a blade as a spreading tool.
(a, b, c) $v_\mathrm{T}$ = 10\,mm/s and $Bo_{\mathrm{50}}$ = 0, 4, 15,
(d, e, f) $v_\mathrm{T}$ = 50\,mm/s and $Bo_{\mathrm{50}}$ = 0, 4, 15, and
(g, h, i) $v_\mathrm{T}$ = 100\,mm/s and $Bo_{\mathrm{50}}$ = 0, 4, 15, respectively. Red dots represent the snapshots in Fig.\,\ref{fig:topViewMFblade}
.}
\label{fig:bladeMF}
\end{figure}
\begin{figure}[H]
\centering
\includegraphics[width=\textwidth]{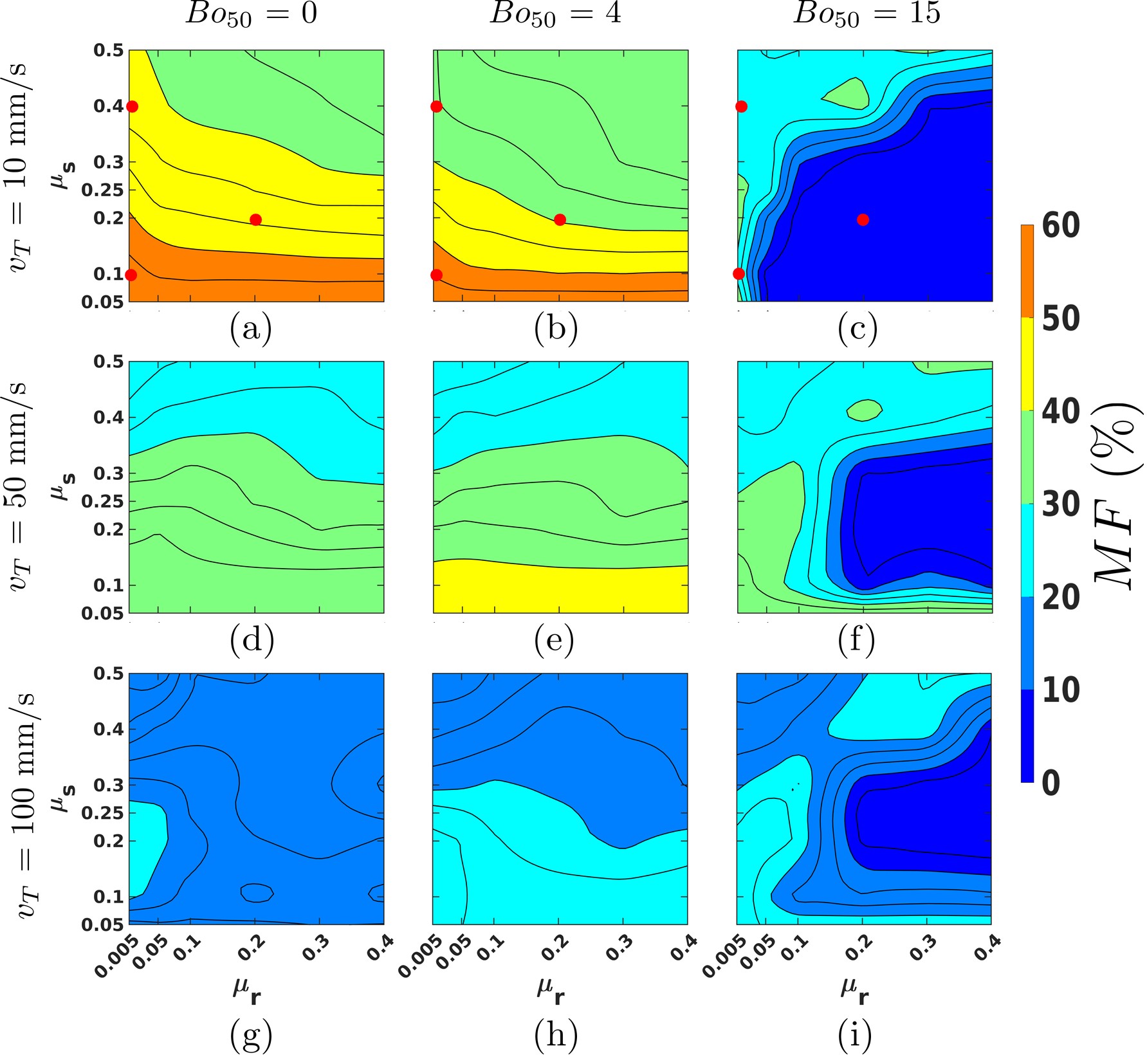}
\caption{Spread layer mass fraction $M\!F$ using a cc wise rotating roller as a spreading tool. 
(a, b, c) $v_\mathrm{T}$ = 10\,mm/s and $Bo_{\mathrm{50}}$ = 0, 4, 15,
(d, e, f) $v_\mathrm{T}$ = 50\,mm/s and $Bo_{\mathrm{50}}$ = 0, 4, 15, and
(g, h, i) $v_\mathrm{T}$ = 100\,mm/s and $Bo_{\mathrm{50}}$ = 0, 4, 15, respectively. Red dots represent the snapshots in Fig.\,\ref{fig:topViewMFroller}
.}
\label{fig:rollerMF}
\end{figure}
%
\input{topViewsMFWith3Regions.tex}
%
\input{topViews4cBladeMF.tex}
%
\input{topViews4cBladeDiffSpeed.tex}
%
\input{topViews4cRollerDiffSpeed.tex}
%
\begin{figure}[H]
\centering
\includegraphics[width=\textwidth]{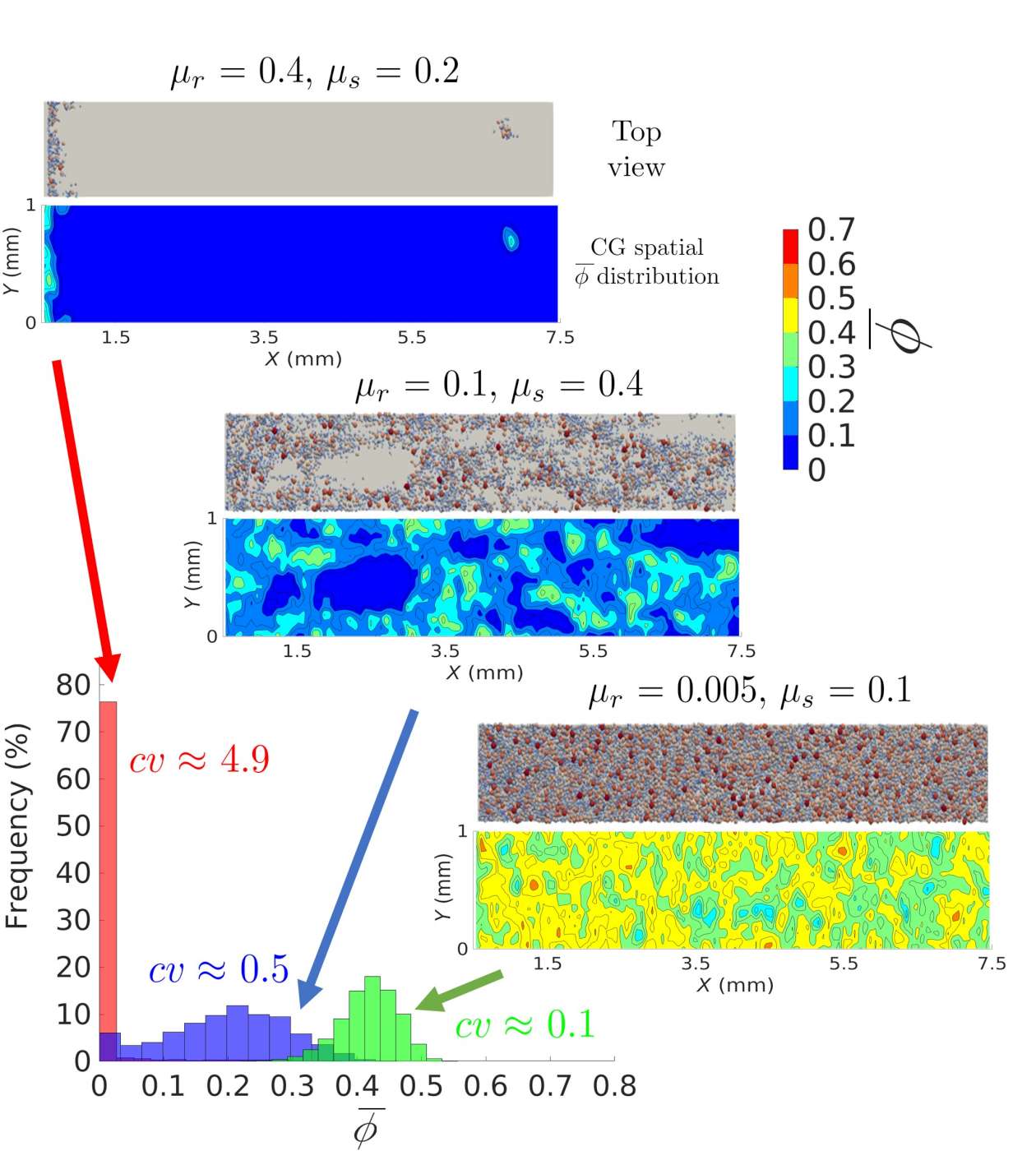}
\caption{The spatial and probability solid volume fraction $\bar{\phi}$  distributions of the spread powder layer with a blade at $v_\mathrm{T}$ = 10\,mm/s, for weakly cohesive particles $Bo_{\mathrm{50}}$ = 4.}
\label{fig:CGmapsBladeVT10NoC}
\end{figure}
\begin{figure}[H]
\centering
\begin{subfigure}[t]{0.45\textwidth}
\subcaption[]{Spreading tool: blade}
\includegraphics[width=\textwidth,angle=0]{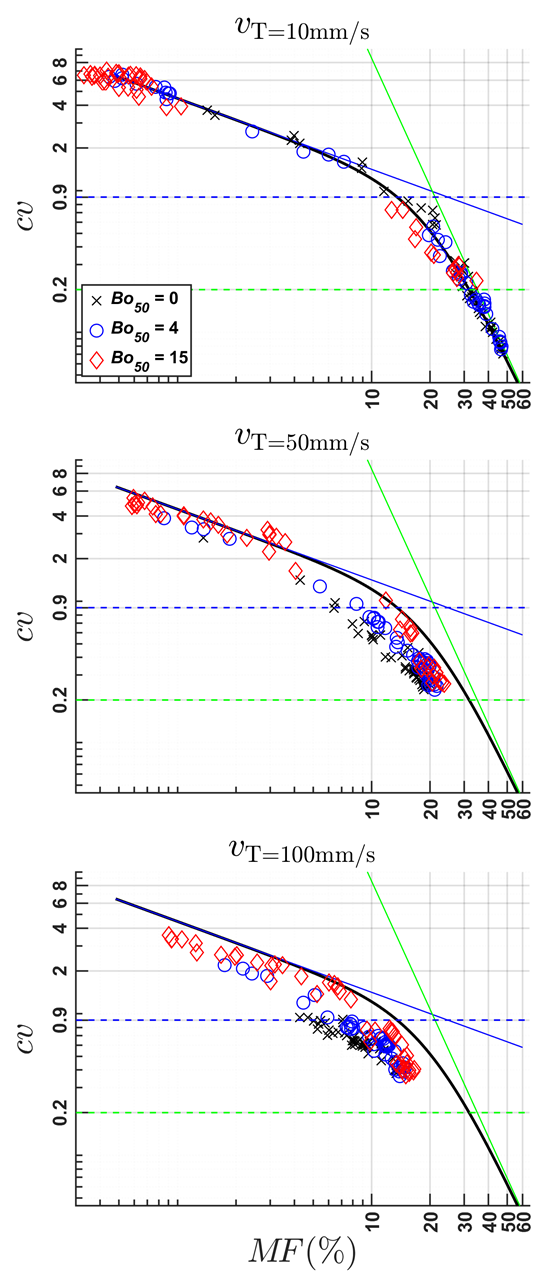}
\label{fig:cvBlade}
\end{subfigure}%
\begin{subfigure}[t]{0.45\textwidth}
\caption[]{Spreading tool: cc rotating roller}
\includegraphics[width=\textwidth,angle=0]{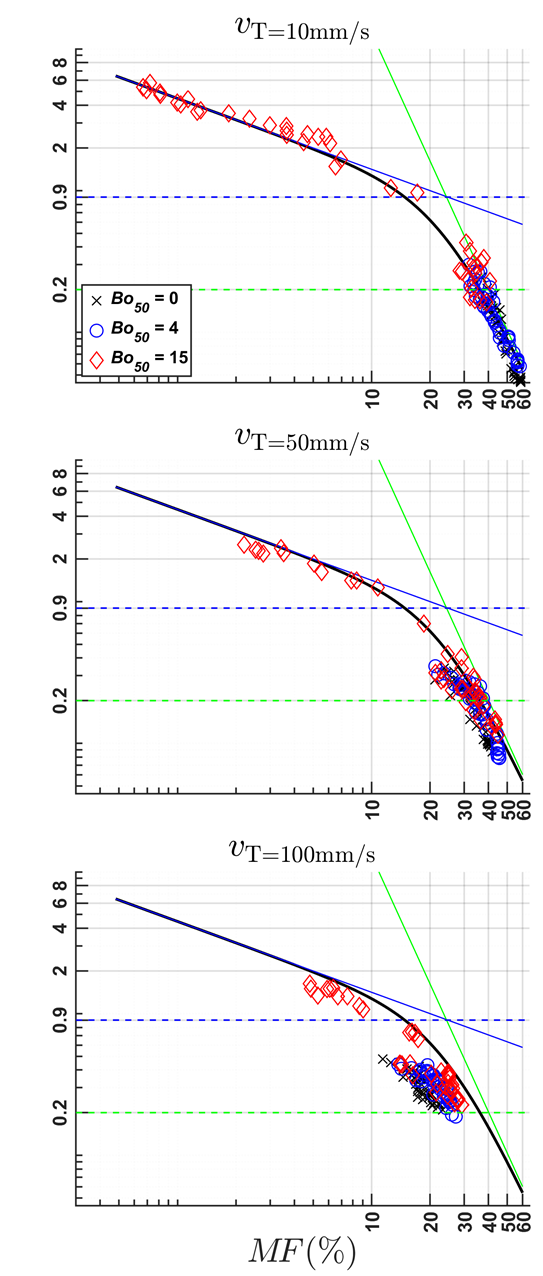}
\label{fig:cvRoller}
\end{subfigure}
\caption{The correlation between $cv$ and $M\!F$. The \textit{dashed green} and \textit{blue horizontal lines} represent the layer uniformity, $cv_u$, and vacancies, $cv_m$, limits, respectively. The \textit{solid black curve} represent the fitting function given by Eq.\ref{eq:fit}, for parameters see main text. 
The \textit{solid blue} and \textit{green lines} represent the power laws in Eq.\ref{eq:fit} i.e. the first and second terms, with power laws $p_1=0.5$ and $p_2=3$, respectively.}
\label{fig:cvAll}
\end{figure}
%
\begin{figure}[H]
\centering
\begin{subfigure}[t]{1\textwidth}
\includegraphics[width=\textwidth,angle=0]{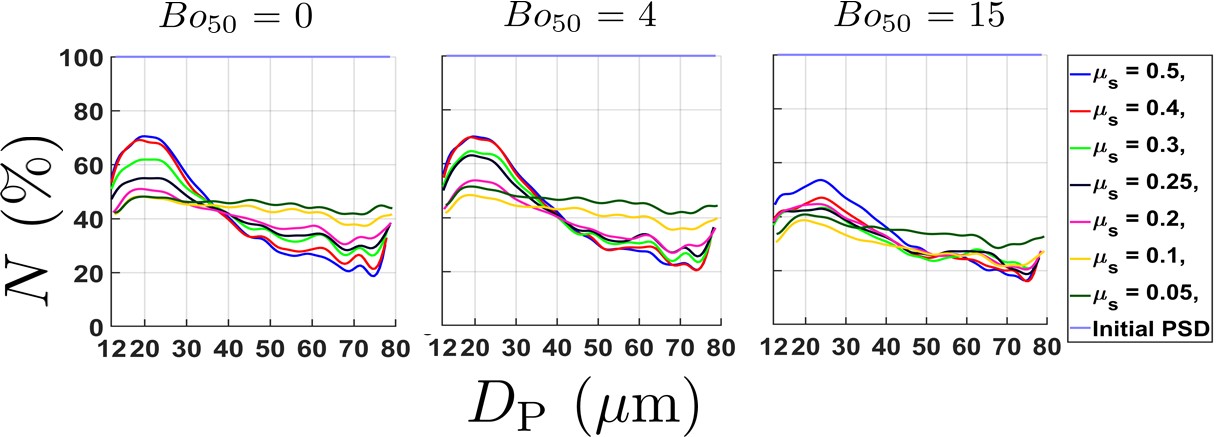}
\caption[]{Fixed $\mu_\mathrm{r} = 0.005$ and varied $\mu_\mathrm{s}$}
\label{fig:bladepsdMuR}
\end{subfigure}%
\hfill
\begin{subfigure}[t]{1\textwidth}
\includegraphics[width=\textwidth,angle=0]{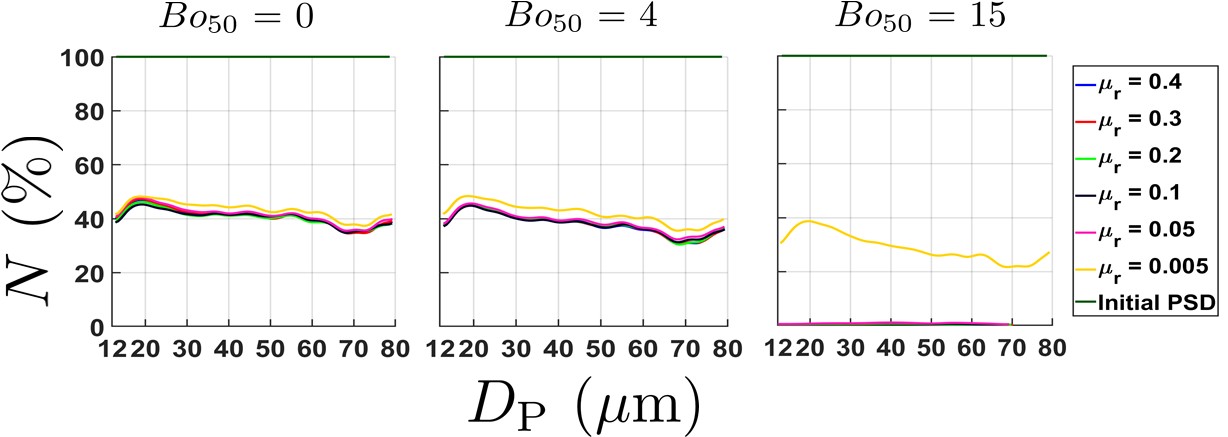}
\caption[]{Fixed $\mu_\mathrm{s} = 0.1$ and varied $\mu_\mathrm{r}$}
\label{fig:bladepsdMuS}
\end{subfigure}
\caption{Spread powder layer particle size distribution using a blade as a spreading tool at $v_\mathrm{T}$ = 10\,mm/s. From left to right, particle cohesiveness $Bo_{\mathrm{50}}$ = 0, 4, 15, respectively.}
\label{fig:bladepsd}
\end{figure}
\begin{figure}[H]
\centering
\begin{subfigure}[t]{1\textwidth}
\includegraphics[width=\textwidth,angle=0]{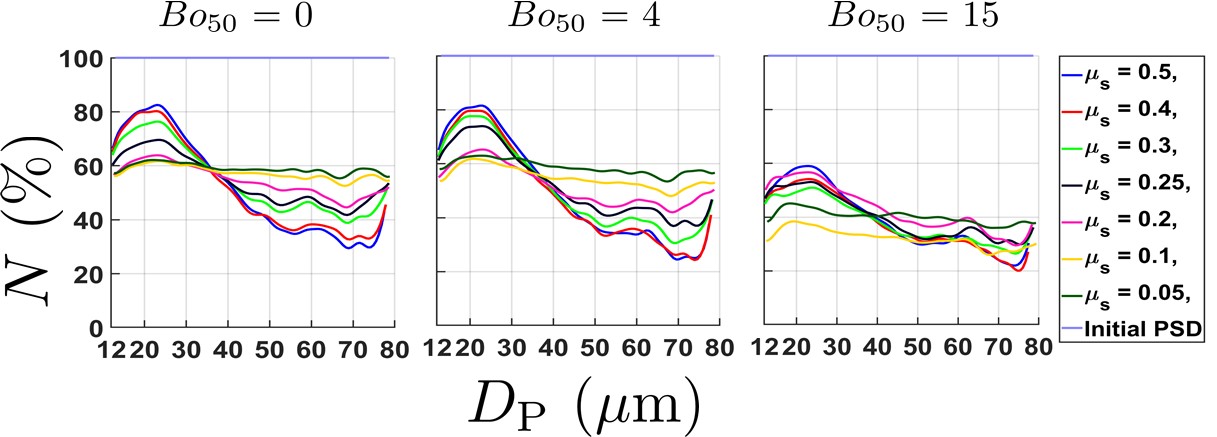}
\caption[]{Fixed $\mu_\mathrm{r} = 0.005$ and varied $\mu_\mathrm{s}$}
\end{subfigure}%
\hfill
\begin{subfigure}[t]{1\textwidth}
\includegraphics[width=\textwidth,angle=0]{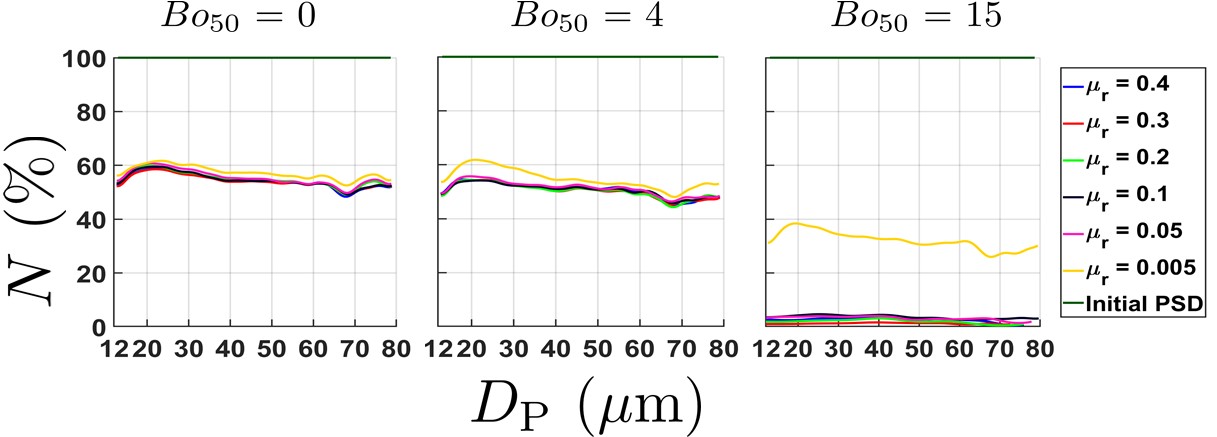}
\caption[]{Fixed $\mu_\mathrm{s} = 0.1$ and varied $\mu_\mathrm{r}$}
\end{subfigure}
\caption{Spread powder layer particle size distribution using a cc rotating roller as a spreading tool at $v_\mathrm{T}$ = 10\,mm/s. From left to right, particle cohesiveness $Bo_{\mathrm{50}}$ = 0, 4, 15, respectively.}
\label{fig:rollerpsd}
\end{figure}
%
\begin{figure}[H]
\centering
\includegraphics[width=\textwidth]{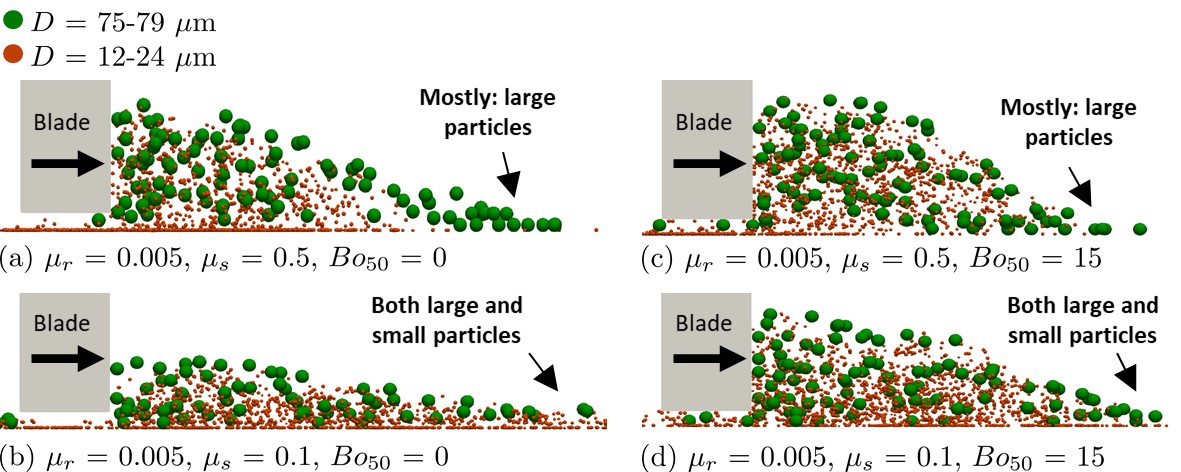}
\caption{Side view of powder heap during the spreading process using a blade at $v_\mathrm{T}$ = 10\,mm/s. Illustrating particles segregation at high and low $\mu_\mathrm{s}$, for non- and strongly cohesive particles.}
\label{fig:bladeSegregation}
\end{figure}
\begin{figure}[H]
\centering
\includegraphics[width=\textwidth]{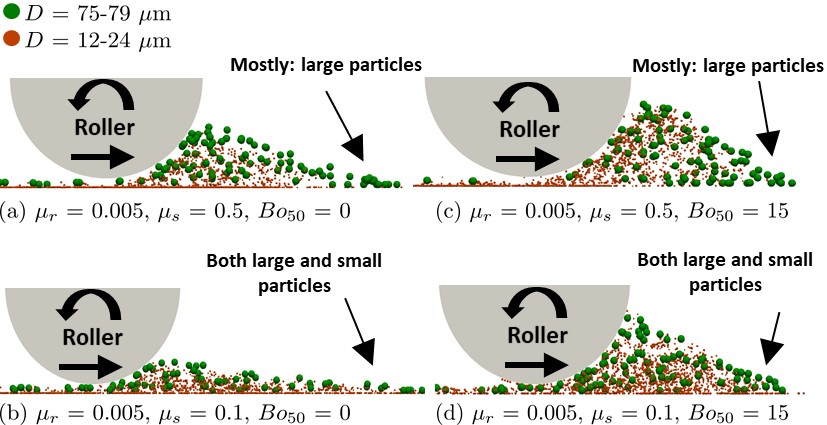}
\caption{Side view of powder heap during the spreading process using a cc rotating roller at $v_\mathrm{T}$ = 10\,mm/s. Illustrating particles segregation at high and low $\mu_\mathrm{s}$, for non- and strongly cohesive particles.}
\label{fig:rollerSegregation}
\end{figure}

%% file: topViewsMFWith3Regions.tex
%
\begin{figure}[H]
\centering
\begin{subfigure}[t]{0.33\textwidth}
\includegraphics[width=\textwidth,angle=0]{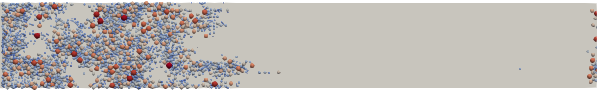}
\caption[]{$\mu_\mathrm{r}$ = 0.2, $\mu_\mathrm{s}$ = 0.4}
\end{subfigure}%
\hfill
\begin{subfigure}[t]{0.33\textwidth}
\includegraphics[width=\textwidth,angle=0]{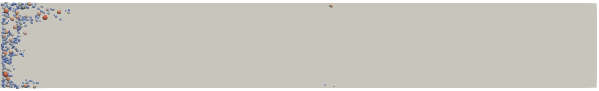}
\caption[]{$\mu_\mathrm{r}$ = 0.2, $\mu_\mathrm{s}$ = 0.4}
\end{subfigure}%
\hfill
\begin{subfigure}[t]{0.33\textwidth}
\includegraphics[width=\textwidth,angle=0]{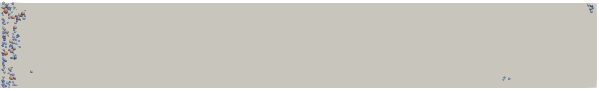}
\caption[]{$\mu_\mathrm{r}$ = 0.2, $\mu_\mathrm{s}$ = 0.4}
\end{subfigure}
\begin{subfigure}[t]{0.33\textwidth}
\includegraphics[width=\textwidth,angle=0]{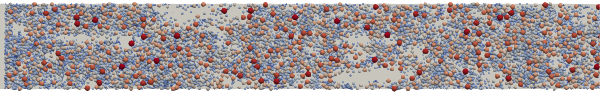}
\caption[]{$\mu_\mathrm{r}$ = 0.1, $\mu_\mathrm{s}$ = 0.3}
\end{subfigure}%
\hfill
\begin{subfigure}[t]{0.33\textwidth}
\includegraphics[width=\textwidth,angle=0]{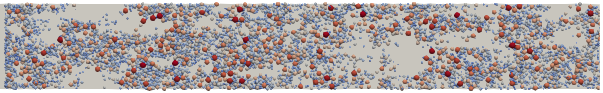}
\caption[]{$\mu_\mathrm{r}$ = 0.1, $\mu_\mathrm{s}$ = 0.3}
\end{subfigure}%
\hfill
\begin{subfigure}[t]{0.33\textwidth}
\includegraphics[width=\textwidth,angle=0]{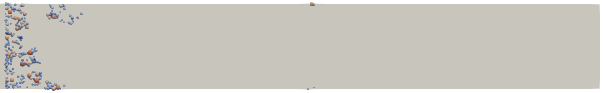}
\caption[]{$\mu_\mathrm{r}$ = 0.1, $\mu_\mathrm{s}$ = 0.3}
\end{subfigure}
\begin{subfigure}[t]{0.33\textwidth}
\includegraphics[width=\textwidth,angle=0]{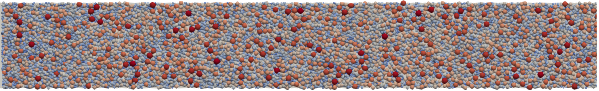}
\caption[]{$\mu_\mathrm{r}$ = 0.005, $\mu_\mathrm{s}$ = 0.1}
\end{subfigure}%
\hfill
\begin{subfigure}[t]{0.33\textwidth}
\includegraphics[width=\textwidth,angle=0]{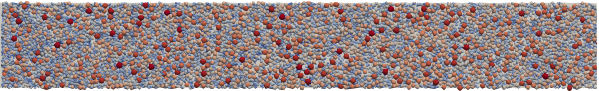}
\caption[]{$\mu_\mathrm{r}$ = 0.005, $\mu_\mathrm{s}$ = 0.1}
\label{fig:topViewBoth3b}
\end{subfigure}%
\hfill
\begin{subfigure}[t]{0.33\textwidth}
\includegraphics[width=\textwidth,angle=0]{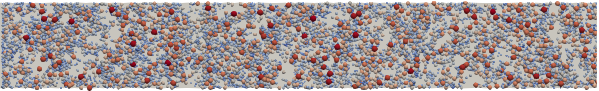}
\caption[]{$\mu_\mathrm{r}$ = 0.005, $\mu_\mathrm{s}$ = 0.1}
\end{subfigure}
\caption{Top view of the spread powder layer using a blade as a spreading tool at $v_\mathrm{T}$\,=\,10\,mm/s. (a, d, g) $Bo_\mathrm{50} = 0$, (b, e, h) $Bo_{50} = 4$ and (c, f, i) $Bo_\mathrm{50} = 15$. Cases:  (a,b,c,f) Empty layers due to global particle drag and particle interlock $M\!F < 10\,\%$, (d, e, i) layers with empty patches due to local particle drag $10\,\% < M\!F < 30\,\%$ and (g, h) relatively good dense layers $M\!F > 30\,\%$.} 
\label{fig:topViewMFblade}
\end{figure}
%
\begin{figure}[H]
\centering
\begin{subfigure}[t]{0.33\textwidth}
\includegraphics[width=\textwidth, angle=0]{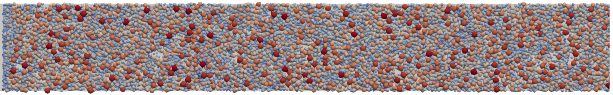}
\caption[]{$\mu_\mathrm{r}$ = 0.2, $\mu_\mathrm{s}$ = 0.2}
\end{subfigure}%
\hfill
\begin{subfigure}[t]{0.33\textwidth}
\includegraphics[width=\textwidth,angle=0]{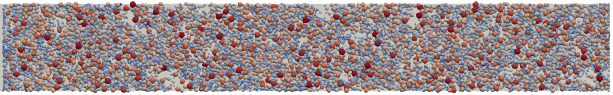}
\caption[]{$\mu_\mathrm{r}$ = 0.2, $\mu_\mathrm{s}$ = 0.2}
\end{subfigure}%
\hfill
\begin{subfigure}[t]{0.33\textwidth}
\includegraphics[width=\textwidth,angle=0]{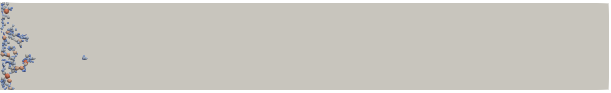}
\caption[]{$\mu_\mathrm{r}$ = 0.2, $\mu_\mathrm{s}$ = 0.2}
\end{subfigure}
\begin{subfigure}[t]{0.33\textwidth}
\includegraphics[width=\textwidth,angle=0]{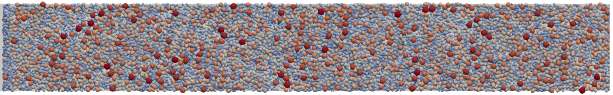}
\caption[]{$\mu_\mathrm{r}$ = 0.005, $\mu_\mathrm{s}$ = 0.4}
\end{subfigure}%
\hfill
\begin{subfigure}[t]{0.33\textwidth}
\includegraphics[width=\textwidth,angle=0]{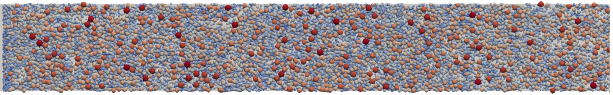}
\caption[]{$\mu_\mathrm{r}$ = 0.005, $\mu_\mathrm{s}$ = 0.4}
\end{subfigure}%
\hfill
\begin{subfigure}[t]{0.33\textwidth}
\includegraphics[width=\textwidth,angle=0]{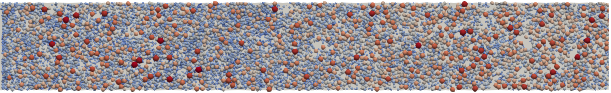}
\caption[]{$\mu_\mathrm{r}$ = 0.005, $\mu_\mathrm{s}$ = 0.4}
\end{subfigure}
\begin{subfigure}[t]{0.33\textwidth}
\includegraphics[width=\textwidth,angle=0]{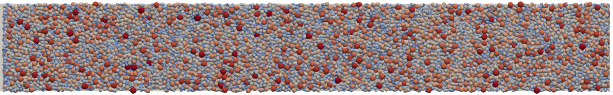}
\caption[]{$\mu_\mathrm{r}$ = 0.005, $\mu_\mathrm{s}$ = 0.1}
\end{subfigure}%
\hfill
\begin{subfigure}[t]{0.33\textwidth}
\includegraphics[width=\textwidth,angle=0]{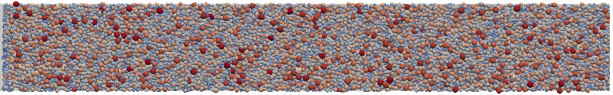}
\caption[]{$\mu_\mathrm{r}$ = 0.005, $\mu_\mathrm{s}$ = 0.1}
\end{subfigure}%
\hfill
\begin{subfigure}[t]{0.33\textwidth}
\includegraphics[width=\textwidth,angle=0]{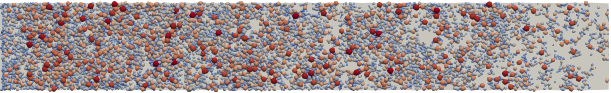}
\caption[]{$\mu_\mathrm{r}$ = 0.005, $\mu_\mathrm{s}$ = 0.1}
\end{subfigure}
\caption{Top view of the spread powder layer using a counter-clock wise rotating as a spreading tool at $v_\mathrm{T}$ = 10\,mm/s. (a, d, g) $Bo_\mathrm{50} = 0$, (b, e, h) $Bo_\mathrm{50} = 4$ and (c, f, i) $Bo_\mathrm{50}\,=\,15$. Cases: (c) Empty layer due to global particle drag and particle interlock $M\!F < 10\,\%$, (b, f, i) layers with empty patches due to local particle drag $10\,\% < M\!F < 30\,\%$ and (a, d, g, e, h) relatively good dense layers $M\!F > 30\,\%$.} 
\label{fig:topViewMFroller}
\end{figure}
%

%% file: topViews4cBladeMF.tex
\begin{figure}[H]
\centering
\begin{subfigure}[t]{0.33\textwidth}
\includegraphics[width=\textwidth,angle=0]{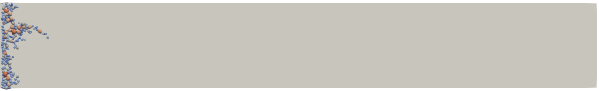}
\caption[]{$\mu_\mathrm{r}$ = 0.05, $\mu_\mathrm{s}$ = 0.1}
\end{subfigure}%
\hfill
\begin{subfigure}[t]{0.33\textwidth}
\includegraphics[width=\textwidth,angle=0]{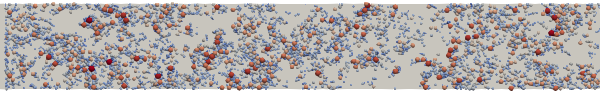}
\caption[]{$\mu_\mathrm{r}$ = 0.05, $\mu_\mathrm{s}$ = 0.3}
\end{subfigure}%
\hfill
\begin{subfigure}[t]{0.33\textwidth}
\includegraphics[width=\textwidth,angle=0]{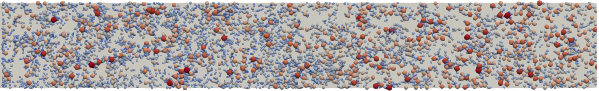}
\caption[]{$\mu_\mathrm{r}$ = 0.05, $\mu_\mathrm{s}$ = 0.4}
\end{subfigure}
\begin{subfigure}[t]{0.33\textwidth}
\includegraphics[width=\textwidth,angle=0]{blade4Cohesion-10Speed-topView41.png}
\caption[]{$\mu_\mathrm{r}$ = 0.005, $\mu_\mathrm{s}$ = 0.1}
\end{subfigure}%
\hfill
\begin{subfigure}[t]{0.33\textwidth}
\includegraphics[width=\textwidth,angle=0]{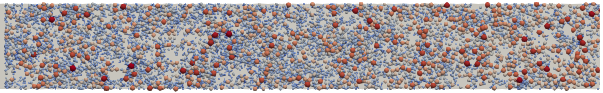}
\caption[]{$\mu_\mathrm{r}$ = 0.005, $\mu_\mathrm{s}$ = 0.3}
\end{subfigure}%
\hfill
\begin{subfigure}[t]{0.33\textwidth}
\includegraphics[width=\textwidth,angle=0]{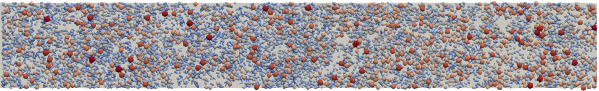}
\caption[]{$\mu_\mathrm{r}$ = 0.005, $\mu_\mathrm{s}$ = 0.4}
\end{subfigure}
\caption{Top view of the spread powder layer for strongly cohesive particles $Bo_{50} = 15$ using a blade as a spreading tool at $v_\mathrm{T}$ = 10\,mm/s, illustrate the effect of $\mu_\mathrm{s}$ and $\mu_\mathrm{r}$.} 
\label{fig:cohesion4BladeEffectMuS}
\end{figure}

%% file: topViews4cBladeDiffSpeed.tex
%
\begin{figure}[H]
\centering
\begin{subfigure}[t]{0.33\textwidth}
\includegraphics[width=\textwidth,angle=0]{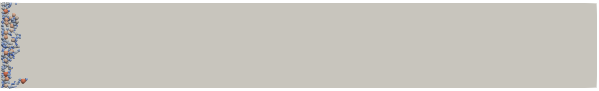}
\caption[]{$v_\mathrm{T}$ = 10\,mm/s}
\end{subfigure}%
\hfill
\begin{subfigure}[t]{0.33\textwidth}
\includegraphics[width=\textwidth,angle=0]{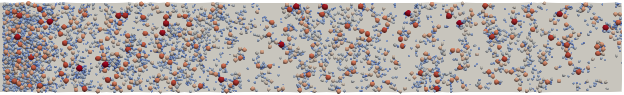}
\caption[]{$v_\mathrm{T}$ = 50\,mm/s}
\end{subfigure}%
\hfill
\begin{subfigure}[t]{0.33\textwidth}
\includegraphics[width=\textwidth,angle=0]{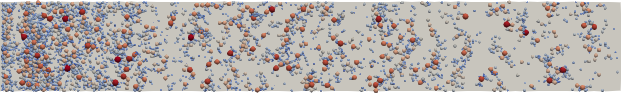}
\caption[]{$v_\mathrm{T}$ = 100\,mm/s}
\end{subfigure}
\caption{Top view of the spread powder layer for strongly cohesive particles $Bo_{50} = 15$, $\mu_\mathrm{r}$ = 0.05 and $\mu_\mathrm{s}$ = 0.05, using a blade as a spreading tool, illustrate the effect of increasing the spreading speed $v_\mathrm{T}$.} 
\label{fig:cohesion4BladeEffectSpeed}
\end{figure}
%

%% file: topViews4cRollerDiffSpeed.tex
%
\begin{figure}[H]
\centering
\begin{subfigure}[t]{0.33\textwidth}
\includegraphics[width=\textwidth,angle=0]{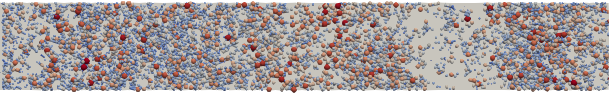}
\caption[]{$\mu_\mathrm{r}$ = 0.4, $\mu_\mathrm{s}$ = 0.5}
\end{subfigure}%
\hfill
\begin{subfigure}[t]{0.33\textwidth}
\includegraphics[width=\textwidth,angle=0]{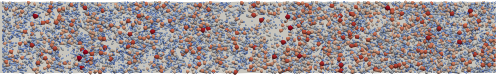}
\caption[]{$\mu_\mathrm{r}$ = 0.4, $\mu_\mathrm{s}$ = 0.5}
\end{subfigure}%
\hfill
\begin{subfigure}[t]{0.33\textwidth}
\includegraphics[width=\textwidth,angle=0]{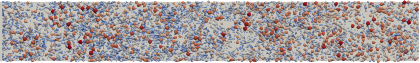}
\caption[]{$\mu_\mathrm{r}$ = 0.4, $\mu_\mathrm{s}$ = 0.5}
\end{subfigure}
\begin{subfigure}[t]{0.33\textwidth}
\includegraphics[width=\textwidth,angle=0]{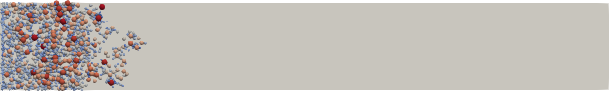}
\caption[]{$\mu_\mathrm{r}$ = 0.4, $\mu_\mathrm{s}$ = 0.05}
\end{subfigure}%
\hfill
\begin{subfigure}[t]{0.33\textwidth}
\includegraphics[width=\textwidth,angle=0]{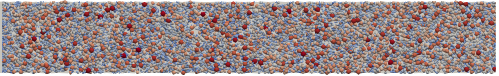}
\caption[]{$\mu_\mathrm{r}$ = 0.4, $\mu_\mathrm{s}$ = 0.05}
\end{subfigure}%
\hfill
\begin{subfigure}[t]{0.33\textwidth}
\includegraphics[width=\textwidth,angle=0]{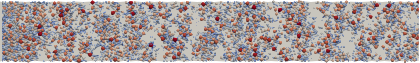}
\caption[]{$\mu_\mathrm{r}$ = 0.4, $\mu_\mathrm{s}$ = 0.05}
\end{subfigure}
\caption{Top view of the spread powder layer for strongly cohesive particles $Bo_\mathrm{50} = 15$, using a roller as a spreading tool at (a, d) $v_\mathrm{T}$ = 10\,mm/s (b, e) $v_\mathrm{T}$ = 50\,mm/s (c, f) $v_\mathrm{T}$ = 100\,mm/s, illustrate the effect of increasing the spreading speed $v_\mathrm{T}$.} 
\label{fig:cohesion4RollerEffectSpeed}
\end{figure}
%

%% file: main.bbl
\begin{thebibliography}{39}
\expandafter\ifx\csname natexlab\endcsname\relax\def\natexlab#1{#1}\fi
\providecommand{\bibinfo}[2]{#2}
\ifx\xfnm\relax \def\xfnm[#1]{\unskip,\space#1}\fi
\bibitem[{Ngo et~al.(2018)Ngo, Kashani, Imbalzano, Nguyen, and Hui}]{intro1}
\bibinfo{author}{T.~D. Ngo}, \bibinfo{author}{A.~Kashani},
  \bibinfo{author}{G.~Imbalzano}, \bibinfo{author}{K.~T. Nguyen},
  \bibinfo{author}{D.~Hui},
\newblock \bibinfo{title}{Additive manufacturing (3d printing): A review of
  materials, methods, applications and challenges},
\newblock \bibinfo{journal}{Composites Part B: Engineering}
  \bibinfo{volume}{143} (\bibinfo{year}{2018}) \bibinfo{pages}{172 -- 196}.
\bibitem[{Campbell et~al.(2012)Campbell, Bourell, and Gibson}]{campbell2012}
\bibinfo{author}{I.~Campbell}, \bibinfo{author}{D.~Bourell},
  \bibinfo{author}{I.~Gibson},
\newblock \bibinfo{title}{Additive manufacturing: rapid prototyping comes of
  age},
\newblock \bibinfo{journal}{Rapid prototyping journal}  (\bibinfo{year}{2012}).
\bibitem[{Sames et~al.(2016)Sames, List, Pannala, Dehoff, and Babu}]{sames2016}
\bibinfo{author}{W.~J. Sames}, \bibinfo{author}{F.~List},
  \bibinfo{author}{S.~Pannala}, \bibinfo{author}{R.~R. Dehoff},
  \bibinfo{author}{S.~S. Babu},
\newblock \bibinfo{title}{The metallurgy and processing science of metal
  additive manufacturing},
\newblock \bibinfo{journal}{International Materials Reviews}
  \bibinfo{volume}{61} (\bibinfo{year}{2016}) \bibinfo{pages}{315--360}.
\bibitem[{Kruth et~al.(2007)Kruth, Levy, Klocke, and Childs}]{Kruth2007}
\bibinfo{author}{J.~P. Kruth}, \bibinfo{author}{G.~Levy},
  \bibinfo{author}{F.~Klocke}, \bibinfo{author}{T.~H. Childs},
\newblock \bibinfo{title}{{Consolidation phenomena in laser and powder-bed
  based layered manufacturing}},
\newblock \bibinfo{journal}{CIRP Annals - Manufacturing Technology}
  \bibinfo{volume}{56} (\bibinfo{year}{2007}) \bibinfo{pages}{730--759}.
\bibitem[{Tan et~al.(2017)Tan, Wong, and Dalgarno}]{TAN2017}
\bibinfo{author}{J.~H. Tan}, \bibinfo{author}{W.~L.~E. Wong},
  \bibinfo{author}{K.~W. Dalgarno},
\newblock \bibinfo{title}{An overview of powder granulometry on feedstock and
  part performance in the selective laser melting process},
\newblock \bibinfo{journal}{Additive Manufacturing} \bibinfo{volume}{18}
  (\bibinfo{year}{2017}) \bibinfo{pages}{228 -- 255}.
\bibitem[{King(2017)}]{report}
\bibinfo{author}{W.~King}, \bibinfo{title}{Modeling of powder dynamics in metal
  additive manufacturing: Final powder dynamics meeting report},
  \bibinfo{year}{2017}.
\bibitem[{Hebert(2016)}]{hebert2016}
\bibinfo{author}{R.~J. Hebert},
\newblock \bibinfo{title}{metallurgical aspects of powder bed metal additive
  manufacturing},
\newblock \bibinfo{journal}{Journal of Materials Science} \bibinfo{volume}{51}
  (\bibinfo{year}{2016}) \bibinfo{pages}{1165--1175}.
\bibitem[{Vock et~al.(2019)Vock, Kl{\"{o}}den, Kirchner, Wei{\ss}g{\"{a}}rber,
  and Kieback}]{Vock2019}
\bibinfo{author}{S.~Vock}, \bibinfo{author}{B.~Kl{\"{o}}den},
  \bibinfo{author}{A.~Kirchner}, \bibinfo{author}{T.~Wei{\ss}g{\"{a}}rber},
  \bibinfo{author}{B.~Kieback},
\newblock \bibinfo{title}{{Powders for powder bed fusion: a review}},
\newblock \bibinfo{journal}{Progress in Additive Manufacturing}
  (\bibinfo{year}{2019}) \bibinfo{pages}{1--15}.
\bibitem[{Herbold et~al.(2015)Herbold, Walton, and Homel}]{Herbold2015}
\bibinfo{author}{E.~Herbold}, \bibinfo{author}{O.~Walton},
  \bibinfo{author}{M.~Homel}, \bibinfo{title}{Simulation of powder layer
  deposition in additive manufacturing processes using the discrete element
  method}, \bibinfo{type}{Technical Report}, Lawrence Livermore National
  Lab.(LLNL), Livermore, CA (United States), \bibinfo{year}{2015}.
\bibitem[{Mindt et~al.(2016)Mindt, Megahed, Lavery, Holmes, and
  Brown}]{Mindt2016}
\bibinfo{author}{H.~W. Mindt}, \bibinfo{author}{M.~Megahed},
  \bibinfo{author}{N.~P. Lavery}, \bibinfo{author}{M.~A. Holmes},
  \bibinfo{author}{S.~G. Brown},
\newblock \bibinfo{title}{{Powder Bed Layer Characteristics: The Overseen
  First-Order Process Input}},
\newblock \bibinfo{journal}{Metallurgical and Materials Transactions A:
  Physical Metallurgy and Materials Science} \bibinfo{volume}{47}
  (\bibinfo{year}{2016}) \bibinfo{pages}{3811--3822}.
\bibitem[{Parteli and P{\"{o}}schel(2016)}]{Parteli2016}
\bibinfo{author}{E.~J. Parteli}, \bibinfo{author}{T.~P{\"{o}}schel},
\newblock \bibinfo{title}{{Particle-based simulation of powder application in
  additive manufacturing}},
\newblock \bibinfo{journal}{Powder Technology} \bibinfo{volume}{288}
  (\bibinfo{year}{2016}) \bibinfo{pages}{96--102}.
\bibitem[{Haeri et~al.(2017)Haeri, Wang, Ghita, and Sun}]{Haeri2017}
\bibinfo{author}{S.~Haeri}, \bibinfo{author}{Y.~Wang},
  \bibinfo{author}{O.~Ghita}, \bibinfo{author}{J.~Sun},
\newblock \bibinfo{title}{{Discrete element simulation and experimental study
  of powder spreading process in additive manufacturing}},
\newblock \bibinfo{journal}{Powder Technology} \bibinfo{volume}{306}
  (\bibinfo{year}{2017}) \bibinfo{pages}{45--54}.
\bibitem[{Haeri(2017)}]{Haeri2017-2}
\bibinfo{author}{S.~Haeri},
\newblock \bibinfo{title}{{Optimisation of blade type spreaders for powder bed
  preparation in Additive Manufacturing using DEM simulations}},
\newblock \bibinfo{journal}{Powder Technology} \bibinfo{volume}{321}
  (\bibinfo{year}{2017}) \bibinfo{pages}{94--104}.
\bibitem[{Chen et~al.(2017)Chen, Wei, Wen, Li, and Shi}]{Chen2017}
\bibinfo{author}{H.~Chen}, \bibinfo{author}{Q.~Wei}, \bibinfo{author}{S.~Wen},
  \bibinfo{author}{Z.~Li}, \bibinfo{author}{Y.~Shi},
\newblock \bibinfo{title}{{Flow behavior of powder particles in layering
  process of selective laser melting: Numerical modeling and experimental
  verification based on discrete element method}},
\newblock \bibinfo{journal}{International Journal of Machine Tools and
  Manufacture} \bibinfo{volume}{123} (\bibinfo{year}{2017})
  \bibinfo{pages}{146--159}.
\bibitem[{Nan et~al.(2018)Nan, Pasha, Bonakdar, Lopez, Zafar, Nadimi, and
  Ghadiri}]{Nan2018a}
\bibinfo{author}{W.~Nan}, \bibinfo{author}{M.~Pasha},
  \bibinfo{author}{T.~Bonakdar}, \bibinfo{author}{A.~Lopez},
  \bibinfo{author}{U.~Zafar}, \bibinfo{author}{S.~Nadimi},
  \bibinfo{author}{M.~Ghadiri},
\newblock \bibinfo{title}{{Jamming during particle spreading in additive
  manufacturing}},
\newblock \bibinfo{journal}{Powder Technology} \bibinfo{volume}{338}
  (\bibinfo{year}{2018}) \bibinfo{pages}{253--262}.
\bibitem[{Nan and Ghadiri(2019)}]{Nan2018b}
\bibinfo{author}{W.~Nan}, \bibinfo{author}{M.~Ghadiri},
\newblock \bibinfo{title}{{Numerical simulation of powder flow during spreading
  in additive manufacturing}},
\newblock \bibinfo{journal}{Powder Technology} \bibinfo{volume}{342}
  (\bibinfo{year}{2019}) \bibinfo{pages}{801--807}.
\bibitem[{Meier et~al.(2019{\natexlab{a}})Meier, Weissbach, Weinberg, Wall, and
  {John Hart}}]{Meier2019a}
\bibinfo{author}{C.~Meier}, \bibinfo{author}{R.~Weissbach},
  \bibinfo{author}{J.~Weinberg}, \bibinfo{author}{W.~A. Wall},
  \bibinfo{author}{A.~{John Hart}},
\newblock \bibinfo{title}{{Modeling and characterization of cohesion in fine
  metal powders with a focus on additive manufacturing process simulations}},
\newblock \bibinfo{journal}{Powder Technology} \bibinfo{volume}{343}
  (\bibinfo{year}{2019}{\natexlab{a}}) \bibinfo{pages}{855--866}.
\bibitem[{Meier et~al.(2019{\natexlab{b}})Meier, Weissbach, Weinberg, Wall, and
  Hart}]{Meier2019b}
\bibinfo{author}{C.~Meier}, \bibinfo{author}{R.~Weissbach},
  \bibinfo{author}{J.~Weinberg}, \bibinfo{author}{W.~A. Wall},
  \bibinfo{author}{A.~J. Hart},
\newblock \bibinfo{title}{{Critical influences of particle size and adhesion on
  the powder layer uniformity in metal additive manufacturing}},
\newblock \bibinfo{journal}{Journal of Materials Processing Technology}
  \bibinfo{volume}{266} (\bibinfo{year}{2019}{\natexlab{b}})
  \bibinfo{pages}{484--501}.
\bibitem[{Han et~al.(2019)Han, Gu, and Setchi}]{Han2019}
\bibinfo{author}{Q.~Han}, \bibinfo{author}{H.~Gu}, \bibinfo{author}{R.~Setchi},
\newblock \bibinfo{title}{{Discrete element simulation of powder layer
  thickness in laser additive manufacturing}},
\newblock \bibinfo{journal}{Powder Technology} \bibinfo{volume}{352}
  (\bibinfo{year}{2019}) \bibinfo{pages}{91--102}.
\bibitem[{Chen et~al.(2019)Chen, Wei, Zhang, Chen, Shi, and Yan}]{Chen2019}
\bibinfo{author}{H.~Chen}, \bibinfo{author}{Q.~Wei},
  \bibinfo{author}{Y.~Zhang}, \bibinfo{author}{F.~Chen},
  \bibinfo{author}{Y.~Shi}, \bibinfo{author}{W.~Yan},
\newblock \bibinfo{title}{{Powder-spreading mechanisms in powder-bed-based
  additive manufacturing: Experiments and computational modeling}},
\newblock \bibinfo{journal}{Acta Materialia} \bibinfo{volume}{179}
  (\bibinfo{year}{2019}) \bibinfo{pages}{158--171}.
\bibitem[{Fouda and Bayly(2020)}]{Fouda2020}
\bibinfo{author}{Y.~M. Fouda}, \bibinfo{author}{A.~E. Bayly},
\newblock \bibinfo{title}{{A DEM study of powder spreading in additive layer
  manufacturing}},
\newblock \bibinfo{journal}{Granular Matter} \bibinfo{volume}{22}
  (\bibinfo{year}{2020}) \bibinfo{pages}{10}.
\bibitem[{Malekipour and El-Mounayri(2018)}]{malekipour2018}
\bibinfo{author}{E.~Malekipour}, \bibinfo{author}{H.~El-Mounayri},
\newblock \bibinfo{title}{Common defects and contributing parameters in powder
  bed fusion am process and their classification for online monitoring and
  control: a review},
\newblock \bibinfo{journal}{The International Journal of Advanced Manufacturing
  Technology} \bibinfo{volume}{95} (\bibinfo{year}{2018})
  \bibinfo{pages}{527--550}.
\bibitem[{Dowling et~al.(2020)Dowling, Kennedy, O'Shaughnessy, and
  Trimble}]{DOWLING2020}
\bibinfo{author}{L.~Dowling}, \bibinfo{author}{J.~Kennedy},
  \bibinfo{author}{S.~O'Shaughnessy}, \bibinfo{author}{D.~Trimble},
\newblock \bibinfo{title}{A review of critical repeatability and
  reproducibility issues in powder bed fusion},
\newblock \bibinfo{journal}{Materials \& Design} \bibinfo{volume}{186}
  (\bibinfo{year}{2020}) \bibinfo{pages}{108346}.
\bibitem[{Andreotti et~al.(2013)Andreotti, Forterre, and Pouliquen}]{GMBook}
\bibinfo{author}{B.~Andreotti}, \bibinfo{author}{Y.~Forterre},
  \bibinfo{author}{O.~Pouliquen}, \bibinfo{title}{Granular media: between fluid
  and solid}, \bibinfo{publisher}{Cambridge University Press},
  \bibinfo{year}{2013}.
\bibitem[{Wensrich and Katterfeld(2012)}]{Wensrich2012}
\bibinfo{author}{C.~M. Wensrich}, \bibinfo{author}{A.~Katterfeld},
\newblock \bibinfo{title}{{Rolling friction as a technique for modelling
  particle shape in DEM}},
\newblock \bibinfo{journal}{Powder Technology} \bibinfo{volume}{217}
  (\bibinfo{year}{2012}) \bibinfo{pages}{409--417}.
\bibitem[{Wensrich et~al.(2014)Wensrich, Katterfeld, and Sugo}]{Wensrich2014}
\bibinfo{author}{C.~M. Wensrich}, \bibinfo{author}{A.~Katterfeld},
  \bibinfo{author}{D.~Sugo},
\newblock \bibinfo{title}{{Characterisation of the effects of particle shape
  using a normalised contact eccentricity}},
\newblock \bibinfo{journal}{Granular Matter} \bibinfo{volume}{16}
  (\bibinfo{year}{2014}) \bibinfo{pages}{327--337}.
\bibitem[{Cundall and Strack(1979)}]{Cundall}
\bibinfo{author}{P.~A. Cundall}, \bibinfo{author}{O.~D.~L. Strack},
\newblock \bibinfo{title}{{A discrete numerical model for granular
  assemblies}},
\newblock \bibinfo{journal}{G{\'{e}}otechnique} \bibinfo{volume}{29}
  (\bibinfo{year}{1979}) \bibinfo{pages}{47--65}.
\bibitem[{Johnson et~al.(1971)Johnson, Kendall, and Roberts}]{cohesion}
\bibinfo{author}{K.~L. Johnson}, \bibinfo{author}{K.~Kendall},
  \bibinfo{author}{A.~D. Roberts},
\newblock \bibinfo{title}{{Surface Energy and the Contact of Elastic Solids}},
\newblock \bibinfo{journal}{Proceedings of the Royal Society A: Mathematical,
  Physical and Engineering Sciences} \bibinfo{volume}{324}
  (\bibinfo{year}{1971}) \bibinfo{pages}{301--313}.
\bibitem[{Luding(2008)}]{luding2008}
\bibinfo{author}{S.~Luding},
\newblock \bibinfo{title}{{Cohesive, frictional powders: Contact models for
  tension}},
\newblock \bibinfo{journal}{Granular Matter} \bibinfo{volume}{10}
  (\bibinfo{year}{2008}) \bibinfo{pages}{235--246}.
\bibitem[{Weinhart et~al.(2012)Weinhart, Thornton, Luding, and
  Bokhove}]{thomasclouser}
\bibinfo{author}{T.~Weinhart}, \bibinfo{author}{A.~R. Thornton},
  \bibinfo{author}{S.~Luding}, \bibinfo{author}{O.~Bokhove},
\newblock \bibinfo{title}{{Closure relations for shallow granular flows from
  particle simulations}},
\newblock \bibinfo{journal}{Granular Matter} \bibinfo{volume}{14}
  (\bibinfo{year}{2012}) \bibinfo{pages}{531--552}.
\bibitem[{Fuchs et~al.(2014)Fuchs, Weinhart, Meyer, Zhuang, Staedler, Jiang,
  and Luding}]{thomasfriction}
\bibinfo{author}{R.~Fuchs}, \bibinfo{author}{T.~Weinhart},
  \bibinfo{author}{J.~Meyer}, \bibinfo{author}{H.~Zhuang},
  \bibinfo{author}{T.~Staedler}, \bibinfo{author}{X.~Jiang},
  \bibinfo{author}{S.~Luding},
\newblock \bibinfo{title}{{Rolling, sliding and torsion of micron-sized silica
  particles: Experimental, numerical and theoretical analysis}},
\newblock \bibinfo{journal}{Granular Matter} \bibinfo{volume}{16}
  (\bibinfo{year}{2014}) \bibinfo{pages}{281--297}.
\bibitem[{Weinhart et~al.(2020)Weinhart, Orefice, Post, van
  Schrojenstein~Lantman, Denissen, Tunuguntla, Tsang, Cheng, Shaheen, Shi,
  Rapino, Grannonio, Losacco, Barbosa, Jing, Naranjo, Roy, den Otter, and
  Thornton}]{mercurydpm}
\bibinfo{author}{T.~Weinhart}, \bibinfo{author}{L.~Orefice},
  \bibinfo{author}{M.~Post}, \bibinfo{author}{M.~P. van Schrojenstein~Lantman},
  \bibinfo{author}{I.~F. Denissen}, \bibinfo{author}{D.~R. Tunuguntla},
  \bibinfo{author}{J.~Tsang}, \bibinfo{author}{H.~Cheng},
  \bibinfo{author}{M.~Y. Shaheen}, \bibinfo{author}{H.~Shi},
  \bibinfo{author}{P.~Rapino}, \bibinfo{author}{E.~Grannonio},
  \bibinfo{author}{N.~Losacco}, \bibinfo{author}{J.~Barbosa},
  \bibinfo{author}{L.~Jing}, \bibinfo{author}{J.~E.~A. Naranjo},
  \bibinfo{author}{S.~Roy}, \bibinfo{author}{W.~K. den Otter},
  \bibinfo{author}{A.~R. Thornton},
\newblock \bibinfo{title}{{Fast, flexible particle simulations — An
  introduction to MercuryDPM}},
\newblock \bibinfo{journal}{Computer Physics Communications}
  \bibinfo{volume}{249} (\bibinfo{year}{2020}) \bibinfo{pages}{107129}.
\bibitem[{Geer et~al.(2018)Geer, Bernhardt-Barry, Garboczi, Whiting, and
  Donmez}]{Geer2018}
\bibinfo{author}{S.~Geer}, \bibinfo{author}{M.~L. Bernhardt-Barry},
  \bibinfo{author}{E.~J. Garboczi}, \bibinfo{author}{J.~Whiting},
  \bibinfo{author}{A.~Donmez},
\newblock \bibinfo{title}{{A more efficient method for calibrating discrete
  element method parameters for simulations of metallic powder used in additive
  manufacturing}},
\newblock \bibinfo{journal}{Granular Matter} \bibinfo{volume}{20}
  (\bibinfo{year}{2018}) \bibinfo{pages}{77}.
\bibitem[{Cleary(2010)}]{CLEARY2010}
\bibinfo{author}{P.~W. Cleary},
\newblock \bibinfo{title}{Dem prediction of industrial and geophysical particle
  flows},
\newblock \bibinfo{journal}{Particuology} \bibinfo{volume}{8}
  (\bibinfo{year}{2010}) \bibinfo{pages}{106 -- 118}.
\bibitem[{Weinhart et~al.(2012)Weinhart, Thornton, Luding, and Bokhove}]{CG}
\bibinfo{author}{T.~Weinhart}, \bibinfo{author}{A.~R. Thornton},
  \bibinfo{author}{S.~Luding}, \bibinfo{author}{O.~Bokhove},
\newblock \bibinfo{title}{{From discrete particles to continuum fields near a
  boundary}},
\newblock \bibinfo{journal}{Granular Matter} \bibinfo{volume}{14}
  (\bibinfo{year}{2012}) \bibinfo{pages}{289--294}.
\bibitem[{Tunuguntla et~al.(2016)Tunuguntla, Thornton, and Weinhart}]{Deepack}
\bibinfo{author}{D.~R. Tunuguntla}, \bibinfo{author}{A.~R. Thornton},
  \bibinfo{author}{T.~Weinhart},
\newblock \bibinfo{title}{{From discrete elements to continuum fields:
  Extension to bidisperse systems}},
\newblock \bibinfo{journal}{Computational Particle Mechanics}
  \bibinfo{volume}{3} (\bibinfo{year}{2016}) \bibinfo{pages}{349--365}.
\bibitem[{Goldhirsch(2010)}]{Goldhirsch}
\bibinfo{author}{I.~Goldhirsch},
\newblock \bibinfo{title}{{Stress, stress asymmetry and couple stress: From
  discrete particles to continuous fields}},
\newblock \bibinfo{journal}{Granular Matter} \bibinfo{volume}{12}
  (\bibinfo{year}{2010}) \bibinfo{pages}{239--252}.
\bibitem[{Foster et~al.(2015)Foster, Reutzel, Nassar, Dickman, and
  Hall}]{Foster2015}
\bibinfo{author}{B.~K. Foster}, \bibinfo{author}{E.~W. Reutzel},
  \bibinfo{author}{A.~R. Nassar}, \bibinfo{author}{C.~J. Dickman},
  \bibinfo{author}{B.~T. Hall},
\newblock \bibinfo{title}{{A brief survey of sensing for metal-based powder bed
  fusion additive manufacturing}},
\newblock in: \bibinfo{editor}{K.~G. Harding}, \bibinfo{editor}{T.~Yoshizawa}
  (Eds.), \bibinfo{booktitle}{Dimensional Optical Metrology and Inspection for
  Practical Applications IV}, volume \bibinfo{volume}{9489},
  \bibinfo{organization}{International Society for Optics and Photonics},
  \bibinfo{publisher}{SPIE}, \bibinfo{year}{2015}, pp. \bibinfo{pages}{40 --
  48}.
\bibitem[{Abdelrahman et~al.(2017)Abdelrahman, Reutzel, Nassar, and
  Starr}]{Abdelrahman2017}
\bibinfo{author}{M.~Abdelrahman}, \bibinfo{author}{E.~W. Reutzel},
  \bibinfo{author}{A.~R. Nassar}, \bibinfo{author}{T.~L. Starr},
\newblock \bibinfo{title}{{Flaw detection in powder bed fusion using optical
  imaging}},
\newblock \bibinfo{journal}{Additive Manufacturing} \bibinfo{volume}{15}
  (\bibinfo{year}{2017}) \bibinfo{pages}{1--11}.

\end{thebibliography}
